\documentclass[11pt]{article}
\usepackage{amsmath, amssymb, amsthm}
\usepackage{graphicx}
\usepackage{fullpage}
\usepackage{setspace}
\usepackage{float}
\usepackage{color}
\usepackage{hyperref}
\usepackage{authblk}
\pdfoutput=1

\title{Algebraic Methods and Computational Strategies for Pseudoinverse-Based MR Image Reconstruction (Pinv-Recon)}
\author[1,2,3]{Kylie Yeung}
\author[4]{Christine Tobler}
\author[5]{Rolf F Schulte}
\author[1]{Benjamin White}
\author[3]{Anthony McIntyre}
\author[6,7]{S\'ebastien Serres}
\author[8]{Peter Morris}
\author[8,9,10]{Dorothee Auer}
\author[2,3]{Fergus V Gleeson}
\author[1,11,$+$]{Damian J Tyler}
\author[1,3,$+,*$]{James T Grist}
\author[5,12,$+$]{Florian Wiesinger}

\affil[1]{Oxford Centre for Clinical Magnetic Resonance (OCMR), University of Oxford, Oxford, OX3 9DU, United Kingdom}
\affil[2]{Department of Oncology, University of Oxford, Oxford, OX3 7DQ, United Kingdom}
\affil[3]{Department of Radiology, Oxford University Hospitals NHS Trust, Oxford, OX3 7LE, United Kingdom}
\affil[4]{MATLAB, The Mathworks, Munich, 81673, Germany}
\affil[5]{GE HealthCare, Munich, 80807, Germany}
\affil[6]{School of Life Sciences, University of Nottingham, Nottingham, NG7 2TQ, United Kingdom}
\affil[7]{The David Greenfield Human Physiology Unit, University of Nottingham, Nottingham, NG7 2UH, United Kingdom}
\affil[8]{Sir Peter Mansfield Imaging Centre, University of Nottingham, Nottingham, NG7 2QX, United Kingdom}
\affil[9]{Mental Health and Clinical Neuroscience, School of Medicine, University of Nottingham, Nottingham, NG7 2TU, United Kingdom}
\affil[10]{NIHR Nottingham Biomedical Research Centre/Nottingham Clinical Research Facilities, QMC, Nottingham, NG7 2UH, United Kingdom}
\affil[11]{Department of Physiology, Anatomy and Genetics, University of Oxford, Oxford, OX1 3PT, United Kingdom}
\affil[12]{Department of Neuroimaging, Institute of Psychiatry, Psychology \& Neuroscience, King's College London, London, SE5 8AF, United Kingdom}
\affil[$^*$]{Corresponding author: james.grist@cardiov.ox.ac.uk}
\affil[$+$]{These authors contributed equally to this work.}

\begin{document}
\flushbottom
\maketitle

\newpage

\begin{abstract}
Image reconstruction in Magnetic Resonance Imaging (MRI) is fundamentally a linear inverse problem, such that the image can be recovered via explicit pseudoinversion of the encoding matrix by solving $\textbf{data} = \textbf{Encode} \times \textbf{image}$ — a method referred to here as Pinv-Recon. While the benefits of this approach were acknowledged in early studies, the field has historically favored fast Fourier transforms (FFT) and iterative techniques due to perceived computational limitations of the pseudoinversion approach.
This work revisits Pinv-Recon in the context of modern hardware, software, and optimized linear algebra routines. We compare various matrix inversion strategies, assess regularization effects, and demonstrate incorporation of advanced encoding physics into a unified reconstruction framework.

While hardware advances have already significantly reduced computation time compared to earlier studies, our work further demonstrates that leveraging Cholesky decomposition leads to a two-order-of-magnitude improvement in computational efficiency over previous Singular Value Decomposition-based implementations. Moreover, we demonstrate the versatility of Pinv-Recon on diverse \textit{in vivo} datasets encompassing a range of encoding schemes, starting with low- to medium-resolution functional and metabolic imaging and extending to high-resolution cases. Our findings establish Pinv-Recon as a versatile and robust reconstruction framework that aligns with the increasing emphasis on open-source and reproducible MRI research.


\end{abstract}

\section{Introduction}

Magnetic resonance imaging (MRI) is a cornerstone of modern medical diagnostics, providing unparalleled soft tissue contrast and functional information. Its widespread adoption, with over 100 million scans performed annually \cite{webbFiveStepsMake2023}, demonstrates its clinical importance. Central to the effectiveness of MRI is the process of image reconstruction \cite{lauterburImageFormationInduced1973,edelmanHistory2014}, which converts the raw k-space data into diagnostic images. MRI reconstruction has traditionally relied on the fast Fourier transform (FFT) to rapidly transform uniformly sampled Cartesian k-space data into images. However, FFT-based methods struggle with non-Cartesian sampling patterns, parallel imaging (multi-coil data), and other complex encoding schemes, necessitating more sophisticated reconstruction algorithms \cite{beattyRapidGriddingReconstruction2005,wrightNonCartesianParallelImaging2014,manMultifrequencyInterpolationFast1997}. As a result, iterative reconstruction techniques and other advanced methods have been developed to address these challenges, improving image quality while managing computational complexity.

From a mathematical perspective, MR image reconstruction can be understood as a linear inverse problem, where the unknown image is encoded into the measured k-space data via an encoding matrix, \(\textbf{Encode}\) \cite{pruessmannAdvancesSensitivityEncoding2001}. Direct inversion of this encoding matrix, while theoretically possible, has historically been deemed computationally prohibitive in terms of both memory requirements and computational time, because the complexity scales with \(O(N^3)\), where $N$ is the number of elements on one side of \textbf{Encode}  \cite{wrightNonCartesianParallelImaging2014,chenReconstructionNMRData1986,kadahAlgebraicReconstructionMagnetic1998,cruzChapter2MRI2022}. This is in contrast to the computational complexity of \(O(N \log N)\) of the FFT. Early attempts, for example Van de Walle \textit{et al.} in 2000 \cite{vandewalleReconstructionMRImages2000}, explored pseudoinverse-based reconstruction but were constrained by the computing resources of that era. Pruessmann’s seminal 2001 paper on sensitivity encoding with arbitrary k-space trajectories also illustrated the challenges of direct matrix inversion on early-2000s hardware and advocated for more efficient iterative solutions \cite{pruessmannAdvancesSensitivityEncoding2001}. Since then, few studies have revisited the computational burden of direct matrix inversion in MRI.

Recent advances in computational hardware and software have significantly altered the landscape of numerical linear algebra computations. The transition to 64-bit operating systems in the early 2000s enabled the handling of very large arrays (exceeding the 4~GB memory limit of 32-bit systems) \cite{wangComputerMemoryStorage2021}, and compatible software packages have emerged in the subsequent years. Progress in multi-threaded central processing units (CPUs), parallel computing platforms, powerful graphical processing units (GPUs), and tensor processing units (TPUs) have driven the rise in artificial intelligence and AI, which in turn have prompted further developments in these areas. Combined with highly optimized numerical linear algebra libraries, these have made large-scale matrix operations, such as direct pseudoinversion of the encoding matrix, increasingly practical for MR image reconstruction applications. 

In this work, we investigate algebraic methods and computational strategies for direct pseudoinversion of the MR encoding matrix, an approach we term `Pinv-Recon'. By leveraging modern computational resources, we demonstrate that Pinv-Recon can be a viable alternative to conventional reconstruction methods, especially for low- to medium-resolution imaging applications. This approach offers several benefits, including simplicity of implementation, inherently interpretable regularization, the ability to account for multiple intertwined encoding mechanisms and distortions in a single framework, and support for computing relevant image reconstruction metrics, such as the spatial response function (SRF) and the noise matrix.

The remainder of this paper is organized as follows. \textit{Algebraic Background} covers potential matrix decomposition methods for directly computing the pseudoinverse of the encoding matrix for MR image reconstruction, expanding beyond the previously suggested method of singular value decomposition (SVD). \textit{Computational Evaluation and Numerical Simulations} evaluates Pinv-Recon in terms of computational feasibility, the impact of different regularization approaches, and its versatility in incorporating various encoding mechanisms and distortions. \textit{Exemplary Applications} include Pinv-Recon of a range of 2D and 3D imaging experiments, including hyperpolarized Carbon-13  brain, hyperpolarized Xenon-129 lung, proton knee, proton brain (the ISMRM CG-SENSE challenge dataset \cite{maierCGSENSERevisitedResults2021}), and proton abdominal imaging. The final section provides a discussion of the advantages, limitations, and potential future directions of Pinv-Recon.

\section{Algebraic Background}

Inspired by the notation introduced by Pruessmann \textit{et al.} in \textit{Encoding and reconstruction in parallel MRI
}\cite{pruessmannEncodingReconstructionParallel2006}, we first formalize the MR image reconstruction problem and then outline possible solutions, including various methods for pseudoinversion of the encoding matrix. We use descriptive variable names in equations for clarity and readability of the algebraic formulation. MRI can be modeled as a linear forward mapping, or ``encoding", of the unknown image onto the raw k-space data, mediated by an encoding matrix, \(\textbf{Encode}\):

\begin{equation}
\textbf{data} = \textbf{Encode} \times \textbf{image},
\label{eq:data_encode_image}
\end{equation}

where both \(\textbf{data}\) and \(\textbf{image}\) have been flattened into 1D vectors. \(\textbf{Encode}\) is a dense, complex matrix that relates each measured data point to all unknown image voxels. In its simplest form, \textbf{Encode} considers only basic gradient encoding (Fourier encoding) and reduces to a Fourier matrix, with each element written as:

\begin{equation}
    \textbf{Encode}_{i_k,i_r}=\mathrm{exp}(j \cdot 2\pi \cdot \textbf{k}_{i_k} \cdot \textbf{r}_{i_r})
    \label{eq:gradient_encode}
\end{equation}

where \(j = \sqrt{-1}\) is the imaginary unit, \(\mathbf{k}_{i_k}\) is the \(i_k\)th k-space sampling point, and \(\mathbf{r}_{i_r}\) is the \(i_r\)th spatial location (voxel position). The k-space points, and spatial locations, can both be discretized to arbitrary grids, including Cartesian, non-Cartesian, uniform, non-uniform, overdiscretized, and masked ones (See Supplementary Material 1).

MRI reconstruction can be considered the inverse problem of this forward mapping, i.e., solving for the unknown \textbf{image} given the measured \textbf{data} and the known \(\textbf{Encode}\) matrix that encapsulates all the encoding physics and geometry. Image reconstruction can be performed by first explicitly inverting \textbf{Encode} to find the reconstruction matrix (\(\textbf{Recon})\) and then applying it to the \textbf{data} by a single matrix-vector multiplication:

\begin{equation}
\textbf{image} = \textbf{Recon} \times \textbf{data}, \qquad \text{where } \textbf{Recon} = \textbf{Encode}^{+},
\end{equation}

This assumes that \textbf{Encode} is  full rank and thus invertible, or has been regularized so that a pseudoinverse (i.e. Moore-Penrose pseudoinverse, denoted by $^{+}$) exists. Alternatively, the \textbf{image} can be obtained without explicitly forming \(\textbf{Recon}\), for example via iterative algorithms.

The following sections highlight several commonly used matrix pseudoinversion methods and linear system solvers. Among these, only truncated SVD-based pseudoinversion and iterative solvers have been commonly considered in the context of MR reconstruction prior to this work \cite{chenReconstructionNMRData1986,kadahAlgebraicReconstructionMagnetic1998,vandewalleReconstructionMRImages2000,kirchnerReductionVoxelBleeding2015}. 

\subsection{Direct Inversion Methods}
The true inverse (denoted by $^{-1}$) of a matrix exists only if the matrix is square and of full rank. However, \(\textbf{Encode}\) is typically rectangular (total number of measurements \(N_k\) by total number of unknowns \(N_r\)) in most MR applications. Direct inversion methods decompose the full rectangular encoding matrix to compute the Moore–Penrose pseudoinverse, which is a generalized solution for inverting such rectangular matrices \cite{barataMoorePenrosePseudoinverse2012}. We highlight two common approaches below.

\textbf{Singular Value Decomposition:} Singular Value Decomposition (SVD) \cite{golubMatrixComputations2013} factorizes and inverts the encoding matrix as:
\begin{equation}
\begin{split}
\textbf{Encode} &= \textbf{U}\,\Sigma\,\textbf{V}^H,
\\
\textbf{Recon} &= \tilde{\textbf{V}}  \,\tilde{\Sigma}^{-1} \tilde{\textbf{U}}^H,
\end{split}
\end{equation}

where \(\textbf{U}\) and \(\textbf{V}\) are unitary matrices (of dimensions \(N_k\times N_k\) and \(N_r\times N_r\), respectively) and \(\Sigma\) is a diagonal matrix containing the singular values \(\sigma_i\) of \textbf{Encode}.  The set of singular values (the SVD spectrum) provides a useful visual impression of the overall conditioning of \textbf{Encode}. $\tilde{\textbf{V}} ,\tilde{\Sigma}, \tilde{\textbf{U}}$ are the truncated versions of each matrix respectively. Truncated SVD (tSVD) regularizes and ensures stability in the inversion by removing small singular values. The ratio of the largest and the remaining smallest singular value in $\tilde{\Sigma}$ is then referred to as the condition number ($\kappa$). Typically, the truncation threshold is set to capture a predetermined percentage of the energy of the original data, such as 95\%. Note that reconstruction matrices with different condition numbers can be realized from the same original untruncated SVD decomposition.

\textbf{QR Decomposition:} QR decomposition \cite{ganderAlgorithmsQRDecomposition1980} factors a matrix into orthogonal and triangular factors and provides the pseudoinverse as:
\begin{equation}
\begin{split}
\textbf{Encode} &= \textbf{Q\,R},
\\
\textbf{Recon} &= \textbf{R}^{-1} \textbf{Q}^H    
\end{split}
\end{equation}

where \(\textbf{Q}\) is an \(N_k \times N_r\) matrix with orthonormal columns and \(\textbf{R}\) is an \(N_r \times N_r\) upper triangular matrix. \textbf{Q} and \textbf{R} may also be truncated for regularization. QR decomposition is typically faster than SVD and maintains good numerical stability, but does not provide the information of the SVD spectrum.

\subsection{Normal-Equation-Based Methods}
Instead of decomposing the full rectangular \textbf{Encode}, the normal equations provide a method to calculate the Moore-Penrose pseudoinverse for both over- and under- determined problems:

\begin{itemize}
  \item \textbf{Overdetermined problems} (\(N_k > N_r\)): where the left-pseudoinverse solution gives 
  \[
  \textbf{Recon} = \textbf{Encode}^{+} = (\textbf{Encode}^H \textbf{Encode})^{-1}\textbf{Encode}^H,
  \]
  providing the least-squares solution for the image (i.e., minimizing \(\|\textbf{Encode}\times\textbf{image} - \textbf{data}\|\)), and
  
  \item \textbf{Underdetermined problems} (\(N_k < N_r\)): where the right-pseudoinverse solution gives 
  \[
  \textbf{Recon} = \textbf{Encode}^{+} = \textbf{Encode}^H (\textbf{Encode}\,\textbf{Encode}^H)^{-1},
  \]
  providing the minimum-norm solution (i.e., 
  $\textbf{image}= \arg\min_{image}||\textbf{image}||^2$ $\;s.t.\;\textbf{data}=\textbf{Encode}\times\textbf{image}$).
\end{itemize}

The two cases can be derived from one another by simple Hermitian transpose operations; for simplicity, we focus on the more
typical overdetermined case going forward. In this case, the $N_r \times N_r$ term $\textbf{Encode}^H \textbf{Encode}$ is referred to as the Gram matrix, which is Hermitian and
positive semi-definite.

Normal-equation-based methods avoid directly inverting the raw rectangular $\mathbf{Encode}$ matrix. Instead, they decompose the smaller, square Gram matrix to derive the pseudoinverse, which is computationally more efficient than decomposing the full rectangular matrix. However, it can be problematic for ill-conditioned problems because the condition number of $\mathbf{Encode}^H \mathbf{Encode}$ is squared compared to that of $\mathbf{Encode}$. This issue can be mitigated by applying appropriate Tikhonov regularization \cite{tikhonovSolutionsIllposedProblems1977,hansenUseLCurveRegularization1993a}, which improves the conditioning of the matrix, similar to singular value truncation. Additionally, Tikhonov regularization helps ensure positive-definiteness, a required property for certain matrix decomposition methods. It is implemented by adding a small $\lambda^2$ to the diagonal of the Gram matrix:

\begin{equation}
\begin{split}
    \textbf{Recon}=\textbf{Encode}^{+} &= (\textbf{Encode}^H \textbf{Encode} + \lambda^2 \textbf{I})^{-1} \textbf{Encode}^H
\\
\textbf{image}_\lambda &= (\textbf{Encode}^H \textbf{Encode} + \lambda^2 \textbf{I})^{-1} \textbf{Encode}^H \textbf{data}.
\label{eq:normal_eq_recon}
\end{split}
\end{equation}

For very large \(\lambda^2\), the solution approaches a simple conjugate-phase reconstruction \cite{pauly_chapter_2005}, whereas smaller \(\lambda^2\) values more fully account for density compensation and deapodization. The choice of \(\lambda^2\) can be guided by principles like Morozov’s discrepancy principle, which suggests selecting \(\lambda^2\) such that the residual \(\|\textbf{Encode}\ \times \textbf{image}_\lambda - \textbf{data}\|\) is on the order of the noise level in the data \cite{Chapter5Tikhonov2012}.

\textbf{Eigenvalue Decomposition:} Eigenvalue decomposition \cite{golubMatrixComputations2013} applies to the Gram matrix and gives \textbf{Recon} using the normal equation:
\begin{equation}
\begin{split}
\textbf{Encode}^H \textbf{Encode} + \lambda^2 \textbf{I} &= \textbf{P}\,\Lambda\,\textbf{P}^{H} , \\
\textbf{Recon}&=\textbf{P}\,\Lambda^{-1} \textbf{P}^{H} \;\textbf{Encode}^H
\end{split}
\end{equation}
where the columns of \(\textbf{P}\) are eigenvectors and \(\Lambda\) is a diagonal matrix of eigenvalues. Note that $\textbf{P}^{-1} =\textbf{P}^{H}$ for a Hermitian matrix. Eigenvalue decomposition is a special case of the SVD applied to a Hermitian matrix (or a symmetric matrix in the real case), thus resulting in only one base \textbf{P}. The eigenvalues of the Gram matrix are equivalent to the singular values of \textbf{Encode} squared. Regularization can be applied by truncating the eigenvalues, by adding $\lambda^2$ to the eigenvalues for Tikhonov regularization, or by combining both.

\textbf{Cholesky Decomposition:} Cholesky decomposition \cite{golubMatrixComputations2013} is known for being one of the most computationally efficient matrix decomposition algorithms, serving as an attractive choice for Pinv-Recon. It is a specialized method for square, Hermitian positive-definite matrices, factorizing the Gram matrix and giving \textbf{Recon} as:
\begin{equation}
\begin{split}
\textbf{Encode}^H \textbf{Encode} + \lambda^2 \textbf{I} &= \textbf{L}\,\textbf{L}^H,
\\
\textbf{Recon} &= (\textbf{L}^{-1})^H \textbf{L}^{-1} \;\textbf{Encode}^{H}
\end{split}
\end{equation}
where \textbf{L} is a lower triangular matrix. The inversion of \textbf{L} is computationally efficient because of its triangular structure.

\subsection{Block-wise Matrix Inversion}
When working with large encoding matrices that exceed available GPU memory, or when parallel computing resources can be exploited, block-wise matrix decomposition can be used \cite{golubMatrixComputations2013}. This involves partitioning \textbf{Encode} or the Gram matrix into smaller blocks, and then applying the algorithm (e.g., factorization, inversion) at the block level, sequentially or in parallel. Block-wise methods are already commonly used internally in linear algebra libraries for performance. In this work, we use the block-wise Cholesky decomposition to overcome GPU memory limitations \cite{chenBlockAlgorithmIts2013}. Note this involves some copying overhead.

\subsection{Direct Solution without explicit inversion}

The matrix inversion methods described above explicitly calculate the pseudoinverse of \textbf{Encode} to obtain the reconstruction operator \textbf{Recon}, and we will refer to them collectively as Pinv-Recon. 
Other approaches exist which directly obtain \textbf{image} without calculating \textbf{Recon}. For example, directly solving equation (\ref{eq:data_encode_image}) with simple matrix division of \textbf{data} by \textbf{Encode}, adding Tikhonov regularization to ensure numerical stability. In practice, this can be implemented using a linear solver (e.g., MATLAB’s backslash operator $A\backslash b$ ), which implicitly performs a matrix
factorization (such as QR or Cholesky-based factorization) to find the least-squares solution without explicitly returning the pseudoinverse.

Iterative algebraic reconstruction techniques, such as conjugate-gradient (CG) methods, may also utilize the explicit encoding matrix and solve for \textbf{image} without finding \textbf{Recon}, and have been widely explored due to their computational efficiency \cite{golubMatrixComputations2013}. For example, it has been explored by Kadah and Hu in  1998 \cite{kadahAlgebraicReconstructionMagnetic1998}, by Wilm \textit{et al.} in 2011 \cite{wilmHigherOrderReconstruction2011}, and by Li \textit{et al.} in 2015 \cite{liAlgebraicReconstructionTechnique2015}, incorporating various encoding mechanisms such as $B_0$ distortion and coil sensitivity which will be further discussed in Section 3.3. We will hereafter refer to this approach as $CG_{Encode}$. Alternative iterative algorithms also exist, for example LSQR \cite{paigeLSQRAlgorithmSparse1982a}.

\subsection{Image Metrics Derived from \textbf{Recon}}

\textbf{Recon}, which is generally not calculated in modern MR image reconstruction algorithms or in direct solution methods such as $CG_{Encode}$, contains useful information for the calculation of relevant image metrics \cite{pruessmannEncodingReconstructionParallel2006,sanchez-gonzalezMinimumnormReconstructionSensitivityencoded2006}. The spatial response function (SRF) measures the proportion of a pixel represented in the reconstructed image. It effectively calculates the forward encoding step followed by the backward projection, and accounts for the fact that each image value does not exclusively reflect the magnetization at the corresponding position, but also exhibits contamination from surrounding signals. Using \textbf{Recon} and \textbf{Encode}, the SRF matrix (\textbf{SRF}) of all image pixels can be calculated by simple matrix-matrix multiplication:

\begin{equation}
    \textbf{SRF} = \textbf{Recon} \times \textbf{Encode}.
\end{equation}

In the optimal scenario, \textbf{SRF} approaches identity. An SRF map can be obtained by reshaping the diagonal elements of \textbf{SRF} to the image size, where a value of 1 indicates the signal from that pixel is fully represented in the final \textbf{image}. \textbf{Recon} also allows simple calculation of the noise matrix \textbf{X}, which reflects the noise propagated during reconstruction:

\begin{equation}
    \textbf{X} = \textbf{Recon} \tilde{\Psi} \textbf{Recon}^H
\end{equation}

where $\tilde{\Psi}$ is the noise covariance matrix.

\section{Computational Evaluation and Numerical Simulations}

\subsection{Computational Time}

\begin{figure}[h!]
    \centering
    \includegraphics[width=0.85\linewidth]{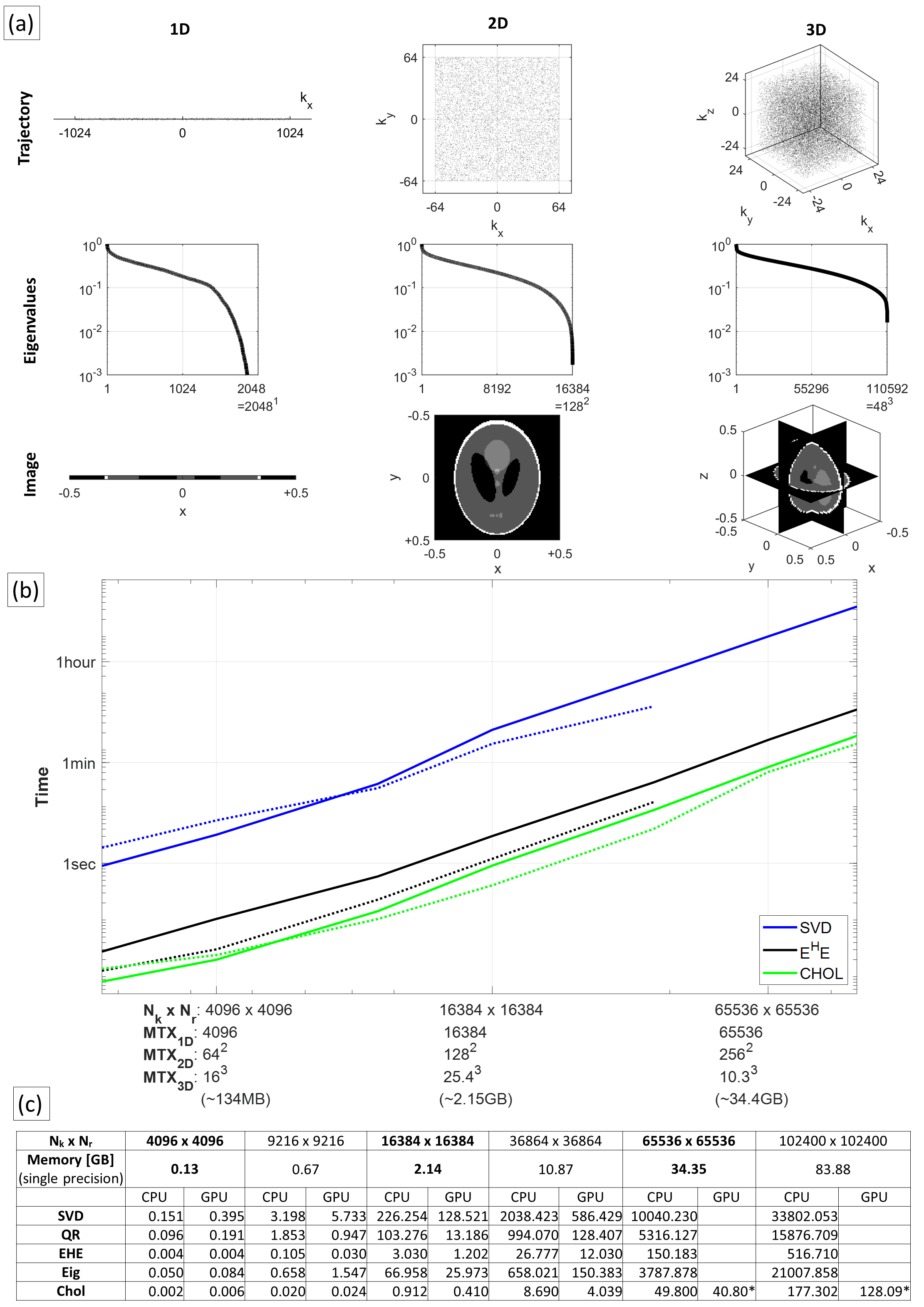}
    \caption{(a) Illustrations of Pinv-Recon using generic 1D, 2D and 3D trajectories. (b) Plot of computational times for SVD, computing \textbf{Encode}$^H$ \textbf{Encode}, and Cholesky decomposition for a range of matrix sizes using CPU only (solid lines) and with GPU acceleration (dotted lines) (c) Corresponding table of computational times also including QR and Eigenvalue decomposition. Asterisks (*) indicate decomposition using block-wise Cholesky decomposition.}
    \label{fig:timing}
\end{figure}

Initial work was undertaken to explore the memory requirements and computation times for decomposing reconstruction matrices using different methods. Random complex single-precision matrices were generated to simulate encoding matrices for 2D image matrix sizes (MTX) 32 $\times$ 32 up to 320 $\times$ 320 (with equivalent MTX for 1D and 3D images noted in the x-axis of Figure \ref{fig:timing}b). Fig. \ref{fig:timing}a illustrates Pinv-Recon of the Shepp-Logan phantom for generic 1D, 2D and 3D random sampling. The matrix decomposition times using SVD, QR, eigenvalue, and Cholesky decomposition in MATLAB (The Mathworks, MA, USA) were measured on a high-performance workstation (Intel\textregistered~Xeon\textregistered~Gold 6448 CPU, 2$\times$32 cores, 1~TB RAM; NVIDIA\textregistered~A100 80~GB GPU), similar to those available on modern MR scanners. The results are summarized in Fig. \ref{fig:timing}b and \ref{fig:timing}c. All computations in MATLAB by default utilized multi-threaded CPU processing across the available CPU cores with the LAPACK (Linear Algebra Package) library \cite{mathworksLAPACKMATLAB}, providing a strong baseline for performance comparisons with GPU acceleration.  GPU acceleration was achieved by assigning \textbf{Encode} to the GPU using \texttt{gpuArray} in MATLAB, the timing of which was not included in the results but can be considered negligible relative to the computation time. Computational times required on a less powerful mobile workstation are also shown in Supplementary Material 2.

The computational time required for a $\sim$2~GB encoding matrix (approximately equivalent to a MTX 128 $\times$ 128 2D image with full Fourier encoding) ranged from under a second to several minutes, depending on the inversion method used. Among the four methods, Cholesky decomposition was fastest, followed by eigenvalue decomposition, QR, and then SVD. This ranking is expected from their algorithmic cost. All methods exhibited scaling approximately proportional to the cube of the matrix dimension (\(O(N^3)\)). Use of the GPU provided additional speedups for the larger matrices (acceleration ranging from 1.7- to 7.8-fold for matrices above 2~GB), though the benefit varied with the decomposition method and matrix size. For very small matrices, the GPU-associated overheads and latencies dominate, and a CPU implementation is more efficient. A block-wise Cholesky algorithm was employed to invert matrices larger than the GPU memory (indicated by asterisks in Fig. \ref{fig:timing}c). While it accelerates computation relative to the CPU, the associated overhead reduces the speedup compared to a non-blockwise approach.

A few additional points are worth noting. For normal-equation-based approaches, the formation of the Gram matrix (computing $\textbf{Encode}^H\textbf{Encode}$) determines the lower bound of the computational time, and therefore is included in Fig. \ref{fig:timing}b and \ref{fig:timing}c for reference. The computational time of iterative solvers is largely determined by factors that influence convergence, such as the matrix’s condition number and the chosen stopping criteria. While these methods can avoid expensive matrix–matrix operations by relying primarily on matrix–vector multiplications, their runtime still grows with the number of iterations required. For conjugate gradient methods that incorporate the FFT, the complexity scales as \(O(N \log N)\), offering significantly greater speed-ups for large matrices compared to smaller ones. The decomposition times for non-Cartesian trajectories (circular coverage of k-space) or masked images (spatial restriction to object coverage) will be shorter than images of equivalent MTX listed in Fig. \ref{fig:timing}c, which assume equal $N_k$ and $N_r$, i.e. full Fourier encoding, for simplicity. For example, a MTX 128 $\times$ 128 image sampled with 15776-point spiral, masked to a circular field-of-view (covering 78\% of image space) results in an encoding matrix of only $N_k \times N_r$ = 15776 $\times$ 12780, and a Gram matrix of 12780 $\times$ 12780, requiring much shorter decomposition time than for $N_k \times N_r$ = 16384 $\times$ 16384 (128$^2$ $\times$ 128$^2$). The present computational evaluation assumes single precision. If double precision (64-bit) were used instead, memory usage and computation time will increase, approximately doubling in many cases; conversely, emerging half-precision (16-bit) or mixed-precision arithmetic could further accelerate these computations at the cost of reduced numerical stability.

\subsection{Regularization}

\begin{figure}[h!]    \centering
    \includegraphics[width=1\linewidth]{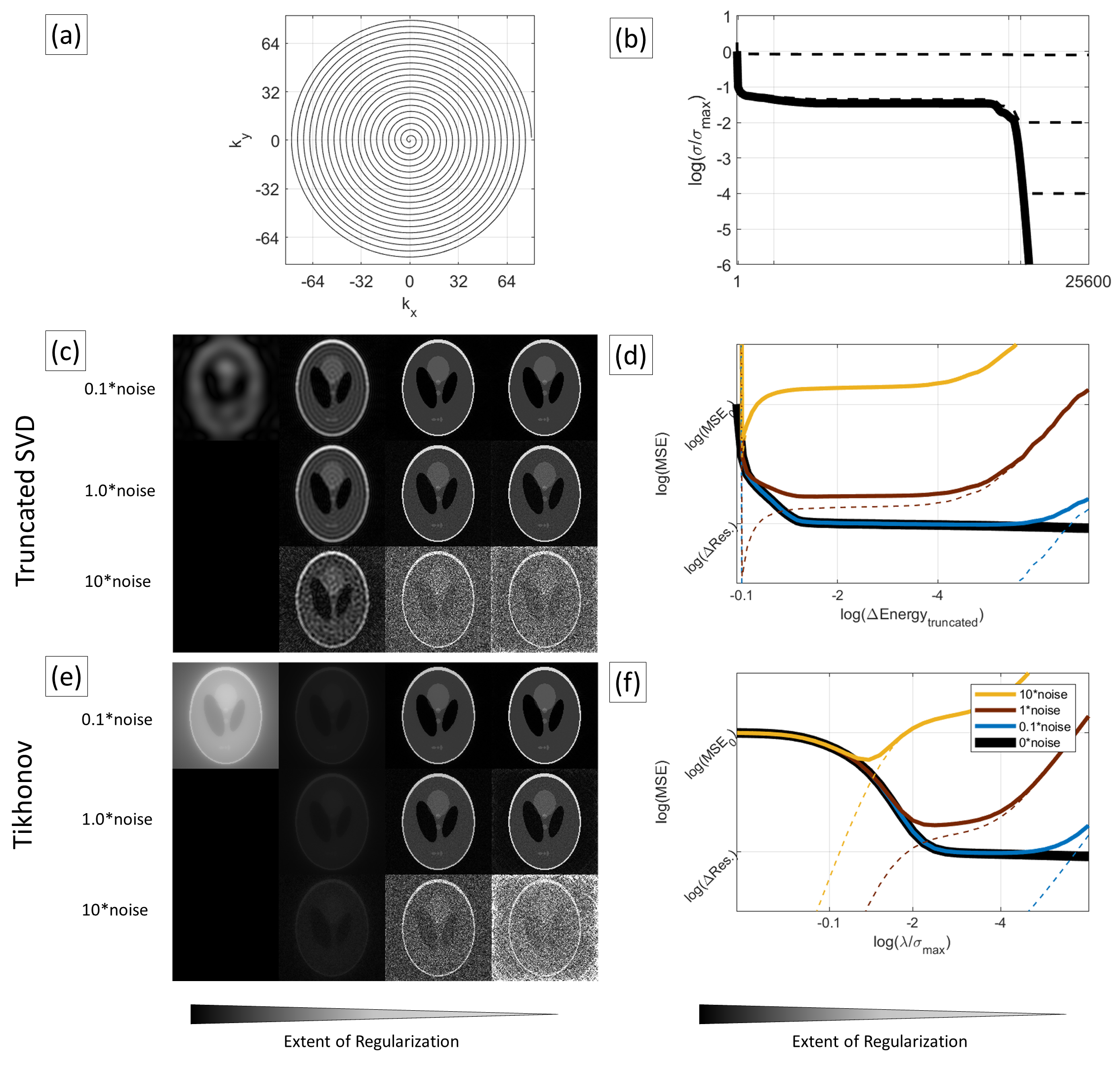}
    \caption{Shepp-Logan Phantom simulations showing the effects of different regularization approaches. (a) The MTX 128 $\times$ 128 spiral trajectory used for simulating the forward encoding and inverse reconstruction. (b) The SVD spectrum of the encoding matrix, with vertical lines showing the truncation of the SVD at different values. The square root of the singular values of \textbf{Encode}$^H$\textbf{Encode} after Tikhonov Regularization at corresponding values are plotted in black dotted lines. (c, e) Shepp-Logan images with different noise levels reconstructed with decreasing regularization from left to right with tSVD and Tikhonov regularization respectively. (d,f) Graphs of MSEs against regularization for varying noise levels, with solid lines showing the MSEs between the original Shepp-Logan image and the reconstructed Shepp-Logan image, and the dotted lines showing the MSEs between purely reconstructing the noise and the ideal Shepp-Logan image. Extent of regularization is expressed for tSVD as the cumulative energy truncated ($\Delta Energy_{truncated}$), and for Tikhonov regularization as the regularization parameter divided by the largest eigenvalue ($\lambda/\sigma_{max}$).}
    \label{fig:regularization}
\end{figure}

Regularization is an essential step in pseudoinverse reconstruction, determining the trade-off between noise amplification and resolution. We compared tSVD versus Tikhonov regularization using a Shepp--Logan phantom simulation. A 128 $\times$ 128 Shepp--Logan phantom was forward-encoded with an 8-arm spiral trajectory (Fig. \ref{fig:regularization}a) and reconstructed using Pinv-Recon under varying noise and regularization settings. The Shepp-Logan images had pixel intensities ranging from 0 to 1, and random complex Gaussian noise was added in k-space at noise levels of 0, 0.1*$noise$, 1*$noise$, and 10*$noise$ (where \texttt{noise=randn(size(Data))+1i*randn(size(Data))/2}). These noise levels were chosen so as to be visible in the image. tSVD reconstructions were performed while varying the energy cutoff (from minimal to near-total truncation), and Tikhonov reconstructions were performed while varying \(\lambda\) (from \(10^{-6}\) to \(10^{5}\) times the largest eigenvalue). 

The log of the SVD spectrum (Fig.~\ref{fig:regularization}b) is shown, normalized to the maximum singular value. Truncation of the SVD at $10^{-0.1}$, $10^{-2}$ and $10^{-4}$ are highlighted with the vertical gray grid lines. The square root of the singular values of \textbf{Encode}$^H$\textbf{Encode} after Tikhonov regularization at corresponding values are plotted in black dotted lines. Fig.~\ref{fig:regularization}c and \ref{fig:regularization}e show reconstructions using tSVD and Tikhonov regularization, respectively, with increasing noise levels (top to bottom) and decreasing regularization (left to right). Fig.~\ref{fig:regularization}d and \ref{fig:regularization}f display mean squared errors (MSE) for different noise levels. The y-axis markers highlight convergence points: $MSE_0$ (error between the Shepp-Logan image and a zero image) and $\Delta Res$ (error between the Shepp-Logan image and an `ideal' reference Shepp-Logan image accounting for the circular k-space field of view).

Insufficient regularization amplifies noise. In low-noise scenarios, minimal truncation suffices for a low MSE. As noise increases, stronger regularization is needed to prevent divergence from the reference image. However, excessive truncation degrades resolution, causing higher MSEs. In SVD truncation, only low-resolution components remain, resulting in a blurry image (Fig. \ref{fig:regularization}c, left). Over-regularizing with Tikhonov regularization (large $\lambda^2$) results in conjugate phase reconstruction, where the reconstructed image converges to a zero image (Fig. \ref{fig:regularization}e, left).

In intermediate regimes, both regularization methods converged to similar reconstruction errors, and all matrix decomposition methods yielded virtually identical results when using the same Tikhonov regularization $\lambda^2$. The normalized MSEs between them are on the order of $10^{-8}$, indicating that the choice of regularization outweighs the choice of decomposition in terms of image reconstruction variability.

\subsection{Generalized Encoding}

We demonstrated the versatility of Pinv-Recon using a numerical Shepp--Logan phantom by incorporating multiple encoding components into \(\textbf{Encode}\). Ideally, \textbf{Encode} can account for all relevant desired encoding mechanisms as well as undesired encoding distortions.  

\textbf{Desired encoding mechanisms} include spatial localization via Fourier encoding (equation (\ref{eq:gradient_encode}), used in the majority of MR imaging), which can be complemented by advanced techniques such as parallel imaging to exploit spatially varying coil sensitivity profiles, and spectral encoding methods like Dixon’s chemical shift encoding using multiple echo times \cite{pruessmannEncodingReconstructionParallel2006,wiesingerIDEALSpiralCSI2012}. These can be mathematically represented as individual factors in the encoding matrix. For example,
\begin{itemize}
    \item \textbf{Coil sensitivity encoding:}
\begin{equation*}
    \textbf{CoilSensitivity}_{i_{Rx}}(\boldsymbol{r}_{i_r})
\end{equation*}

        where \(\textbf{CoilSensitivity}_{i_{Rx}}\) is the spatial sensitivity profile of the \(i_{\mathrm{Rx}}\)th receive coil, and
    \item \textbf{Multi-echo chemical shift encoding:}        
 \begin{equation*}
    \mathrm{exp}({j \cdot \Delta \textbf{CS}_{i_{CS}} \cdot \textbf{TE}_{i_{TE}}}
    \label{eq:desired})
\end{equation*}
        where \(\Delta\textbf{CS}_{i_{\mathrm{CS}}}\) is the chemical shift frequency of the \(i_{\mathrm{CS}}\)th species (with $\textbf{TE}_{i_{TE}}$ the $i_{TE}$th echo time).
\end{itemize}

\textbf{Undesired encoding distortions} inevitably occur in addition to these intended encoding mechanisms, and should be accounted for. Common distortions include:
\begin{itemize}
    \item \textbf{\(B_0\) off-resonance:} 
\begin{equation*}
     \mathrm{exp}({j\cdot \Delta \textbf{B0}_{i_r} \cdot t_{i_k}})
\end{equation*}
    where \(\Delta \textbf{B0}_{i_r}\) is the local field inhomogeneity at voxel \(i_r\), and \(t_{i_k}\) is the time of the \(i_k\)th readout \cite{haskellOffresonanceArtifactCorrection2023},
    \item \textbf{Spatiotemporal gradient imperfections:} 
\begin{equation*}
   \textbf{k}_{i_k} \rightarrow \textbf{k}'_{(i_k,i_r)}
\end{equation*}
    where the true spatiotemporal k-space location $\textbf{k}'_{(i_k,i_r)}$ replaces the nominal k-space coordinates $\textbf{k}_{i_k}$ \cite{duynSimpleCorrectionMethod1998,rahmerRapidAcquisition3D2019,barmetSpatiotemporalMagneticField2008a,sipila2HTransmitReceive2011a}, and

    \item \textbf{Patient motion:}
    \[
    \mathbf{r}_{(i_r,i_k)} \rightarrow Rot(t_{i_k})\,\mathbf{r}_{i_r} + \Delta \mathbf{r}(t_{i_k}),
    \]
    where \(Rot(t_{i_k})\) is a rotation matrix and \(\Delta \mathbf{r}(t_{i_k})\) is a translation vector describing the motion (e.g. due to patient movement) at time \(t_{i_k}\). This simplifies to a constant rigid-body shift if the imaging volume is not centered at the gradient isocenter but does not move during imaging.
\end{itemize}

Each of these factors can be incorporated as a multiplicative term modifying the encoding matrix or as a replacement for nominal k-space or image space coordinates. These encoding components can multiply together to form the net encoding relationship without breaking the linear mapping between \textbf{image} and \textbf{data}, allowing Pinv-Recon to solve for all these effects in one system.

\begin{figure}[h!]    \centering
    \includegraphics[width=\linewidth]{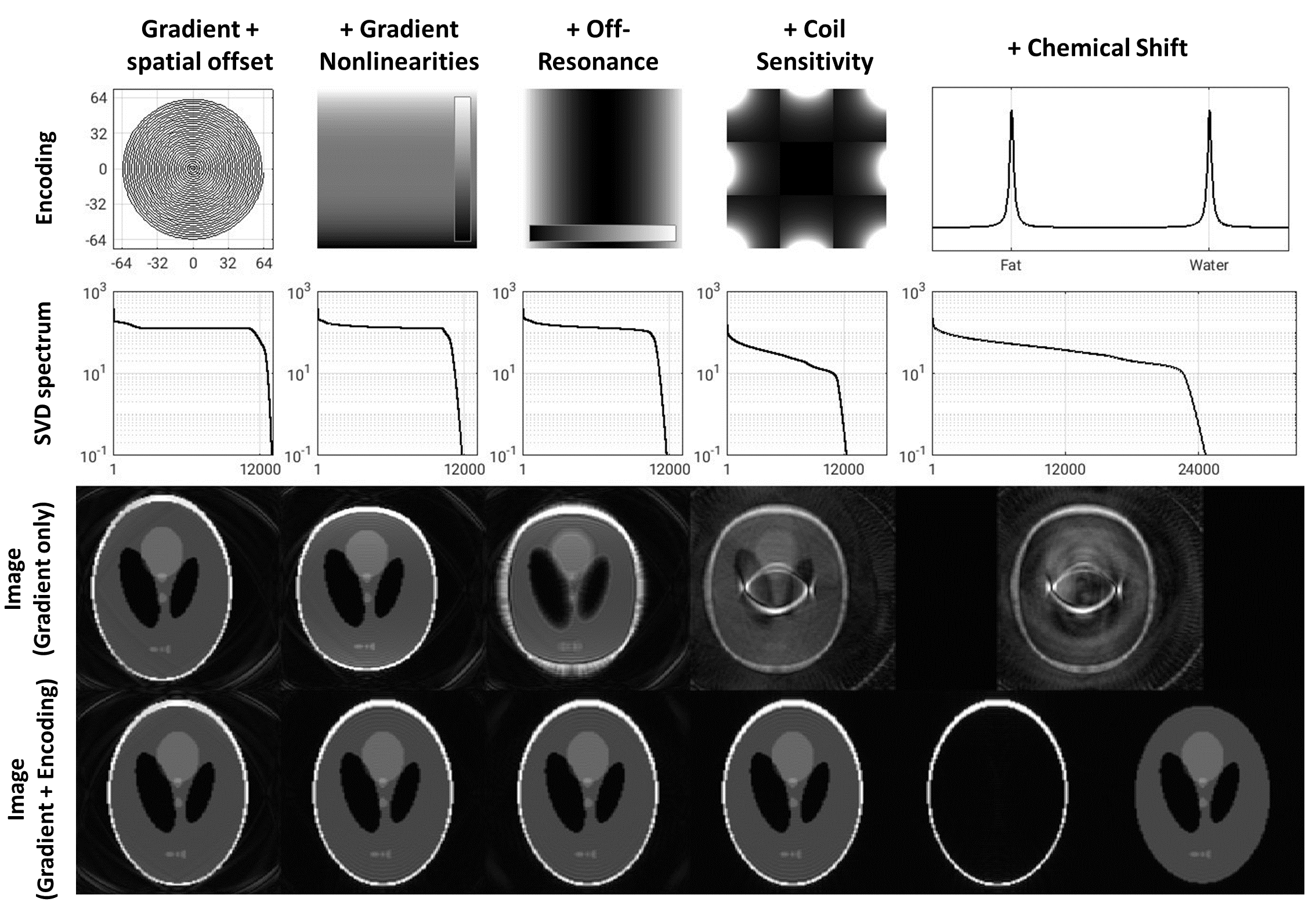}
    \caption{Illustration of additional encoding included in Pinv-Recon as a single unified framework. From top to bottom: encoding mechanism, SVD spectrum of encoding matrix, reconstructed image accounting for gradient encoding only, reconstructed image accounting for gradient and other encoding. From left to right: spiral gradient encoding + spatial off-center positioning, + gradient nonlinearities, + $B_0$ off-resonance, + coil sensitivity encoding (with 2x2 undersampling of the data), + chemical shift separation (of water and fat).}
    \label{fig:generalized-encoding}
\end{figure}

Fig. \ref{fig:generalized-encoding} demonstrates incrementally including all of the above mechanisms and distortions, with the encoding matrix in the final column being:
\[
\begin{split}
\textbf{Encode}_{(i_k,i_{\mathrm{Rx}},i_{\mathrm{TE}}),(i_r,i_{\mathrm{CS}})} \;=\;& 
\underbrace{\exp\!\big(j\,2\pi\,\mathbf{k'}_{(i_k,i_r)}\cdot(\mathbf{r}_{i_r}+\Delta\textbf{r})\big)}_{\text{Gradient}} \;\times\; 
\underbrace{\textbf{CoilSensitivity}_{i_{\mathrm{Rx}}}(\mathbf{r}_{i_r}+\Delta\textbf{r})}_{\text{Coil Sensitivity}} 
\times\; 
\\
&\underbrace{\exp\!\big(j\,\Delta\textbf{CS}_{i_{\mathrm{CS}}}\,\text{TE}_{i_{\mathrm{TE}}}\big)}_{\text{Chemical Shift}} 
\times\; 
\underbrace{\exp\!\big(j\,\Delta \textbf{B0}_{i_r}\,t_{i_k}\big)}_{\text{Off-Resonance}}~.
\end{split}
\]
where the k-space vector and the spatial coordinates were modified as $\textbf{k}_{i_k} \rightarrow \textbf{k}'_{i_k,i_r}$ and $\mathbf{r}_{i_r} \rightarrow \mathbf{r}_{i_r} + \Delta \mathbf{r}$ respectively to include gradient imperfections and patient motion. The data was also undersampled (acceleration factor 2 $\times$ 2) with the inlcusion of the coil sensitivity encoding in the column 4. Row 3 shows reconstruction using \textbf{Encode} accounting for Gradient encoding only, and row 4 shows reconstruction with all relevant encoding accounted for.

In addition to the encoding examples highlighted here, other forms of desired and undesired encoding may also be included \cite{kingSENSEImageQuality2001,maudsleyFourierImagingUsing1986,kartauschSpatialPhaseEncoding2014,sacolickB1MappingBlochSiegert2010}, for example simultaneous multi-slice imaging and spatially varying relaxation decay.

One memory handling strategy for constructing an encoding matrix with coil sensitivity is to perform it in a procedural manner for each coil. Consider a dataset acquired with $N_{Rx}$ coils, each coil with $N_k$ k-space points and $N_r$ image space points. The resulting full $\textbf{Encode}$ matrix is of dimensions ($N_{Rx} \times N_k$) $\times N_r$, and the Gram matrix $\textbf{Encode}^H\textbf{Encode}$ has dimensions $N_r \times N_r$. Rather than constructing and multiplying $\textbf{Encode}$ in its entirety, $\textbf{Encode}_{i_{Rx}}^H\textbf{Encode}_{i_{Rx}}$ can be computed for each coil and added together (equation (\ref{eq:coil_procedural})), significantly reducing computational time.

\begin{equation}
    \begin{split}
    \textbf{Encode}^H\textbf{Encode} &=  \begin{bmatrix} \textbf{Encode}_1^H & \textbf{Encode}_2^H &\cdots&\textbf{Encode}_{N_{Rx}}^H \end{bmatrix}\begin{bmatrix} \textbf{Encode}_1 \\ \textbf{Encode}_2 \\ \cdots \\ \textbf{Encode}_{N_{Rx}} \end{bmatrix}
    \\
    &=\sum_{i_{Rx}=1}^{N_{Rx}}\textbf{Encode}_{i_{Rx}}^H\textbf{Encode}_{i_{Rx}}
    \\
    where  &\; \textbf{Encode}_{i_{Rx}} = \textbf{CoilSensitivity}_{i_{\mathrm{Rx}}}(r_{i_r}) \times \mathrm{exp}(j \cdot 2\pi \cdot \textbf{k}_{i_k} \cdot \textbf{r}_{i_r})
    \end{split}
    \label{eq:coil_procedural}
\end{equation}

Then, instead of preforming a large matrix-matrix multiplication between the inverted Gram matrix and $\textbf{Encode}^H$ (equation (\ref{eq:normal_eq_recon})), $\textbf{Encode}_{i_{Rx}}^H\textbf{data}$ can be computed for each coil individually then summed together, before multiplying with the inverted Gram matrix to obtain the reconstructed image. 

\section{Exemplary Applications}

Several emerging MRI applications feature small to medium matrix sizes and could benefit from non-FFT-based reconstruction, making them good candidates for Pinv-Recon. Examples include hyperpolarized MRI \cite{larsonHyperpolarizedMetabolicMRI2021}, diffusion MRI \cite{kiselevFundamentalsDiffusionMRI2017}, certain functional MRI acquisitions \cite{hernandez-garciaRecentProgressASL2019}, multi-nuclear spectroscopic imaging \cite{weiMultinuclearMagneticResonance2022}, and low-field MRI \cite{arnoldLowfieldMRIClinical2023}. These often have low SNR and use non-Cartesian trajectories or require distortion corrections, which are challenging for traditional FFT-based pipelines but are naturally handled in the explicit encoding matrix formulation. In the following subsections, we demonstrate Pinv-Recon in various experimental scenarios, beginning with smaller image sizes, and then pushing to larger matrix sizes to show the possibilities of Pinv-Recon. Additional examples are shown in Supplementary Materials 3 and 4.

\subsection{Low Resolution 2D Spectral-Spatial Spiral (Hyperpolarized Carbon-13 MRI)}

\begin{figure}[h!]    \centering
    \includegraphics[width=0.8\linewidth]{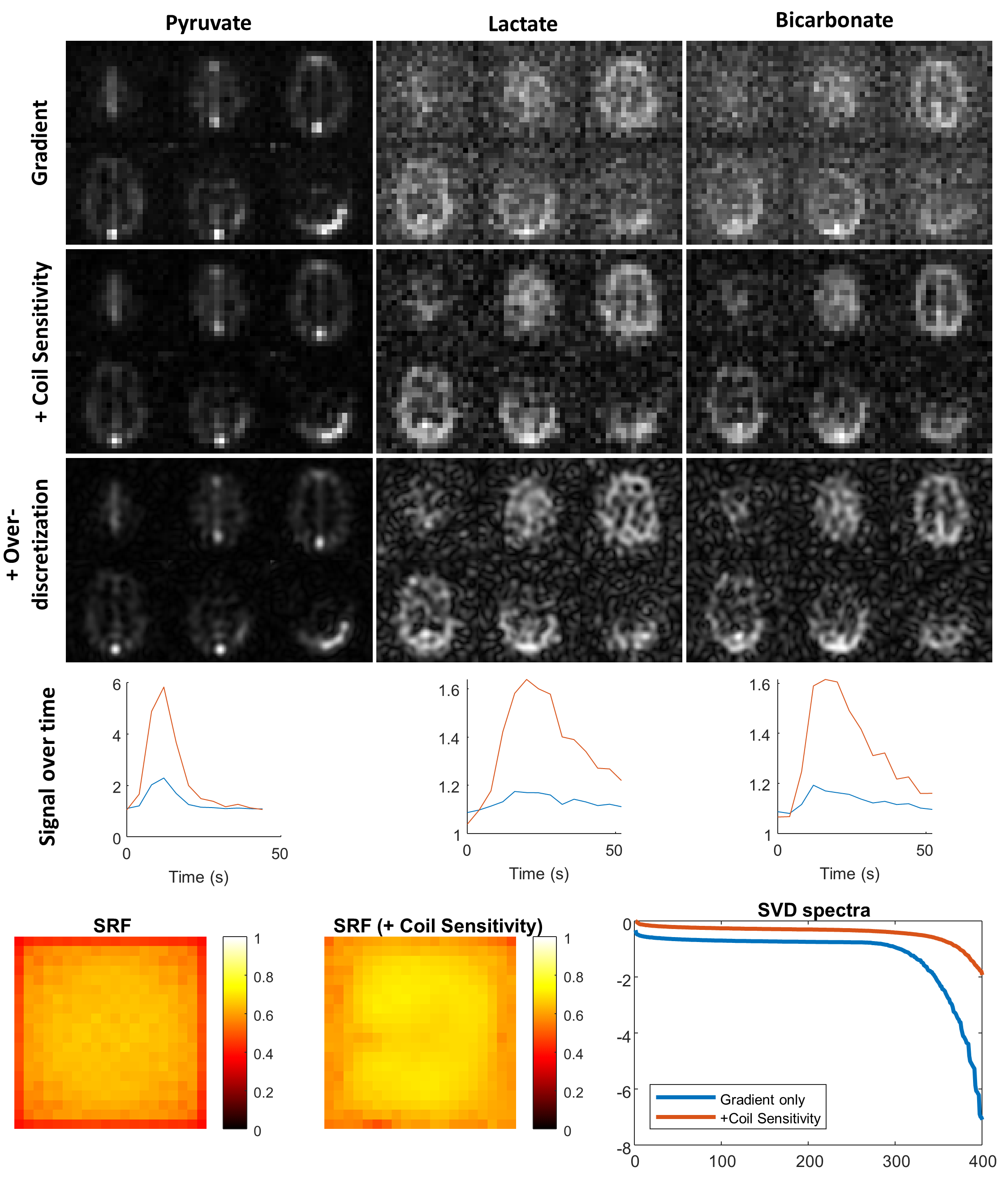}
    \caption{Application of Pinv-Recon on multichannel hyperpolarized carbon-13 data. From left to right: Pyruvate, Lactate, Bicarbonate. From top to bottom: root-sum-of-squares coil combination, Pinv-Recon with coil sensitivity map included, Pinv-Recon with coil sensitivity map included on an overdiscretized spatial grid (MTX 64 × 64), mean SNR for the first two methods (in blue and red respectively), SRFs and SVD spectra from the reconstruction with and without coil sensitivity included respectively. In the first three rows, only six out of eight acquired slices are shown, as there is little signal in the remaining slices.}
    \label{fig:c13}
\end{figure}

As shown in Fig. \ref{fig:c13}, we first applied Pinv-Recon to an 8-channel hyperpolarized \(^{13}\)C metabolic imaging dataset acquired with a 2D spiral trajectory (MTX 20 $\times$ 20) \cite{schulteSaturationrecoveryMetabolicexchangeRate2013}.

Data was reconstructed using (1) Pinv-Recon with gradient encoding only and root-sum-of-squares coil combination, (2) Pinv-Recon with coil sensitivity encoding included, as well as (3) Pinv-Recon with coil sensitivity encoding and overdiscretization to a higher-resolution spatial grid of MTX 64 $\times$ 64. Since the coil sensitivity is spatially-varying but time-invariant, the reconstruction matrix has to be calculated for each of the eight individual slices, but can be used to reconstruct the images across all time points and metabolites in eight matrix-matrix multiplications. Including the coil sensitivity profiles allowed for a more SNR-optimal combination than simple root-sum-of-squares coil combination. Overdiscretized reconstruction can be straightforwardly implemented by defining a finer range of \textbf{r} (equation (\ref{eq:gradient_encode})), which is much simpler compared to over-discretized reconstruction using gridding, often applied by interpolating the data to a finer Cartesian grid \cite{beattyRapidGriddingReconstruction2005,schombergGriddingMethodImage1995}. Pinv-Recon also allowed the calculation of the SRF and noise maps demonstrating that inclusion of the coil sensitivities improves the SRF. Moreover, Pinv-Recon provides inherent and straightforward denoising by regularization, either by truncation of the SVD or by Tikhonov regularization, which is especially useful for SNR-limited techniques such as hyperpolarized carbon-13 imaging \cite{kimDenoisingHyperpolarized13C2021}.

\subsection{Medium Resolution 3D Stack-of-Spirals (Hyperpolarized Xenon-129 MRI)}

\begin{figure}[h!]    \centering
    \includegraphics[width=\linewidth]{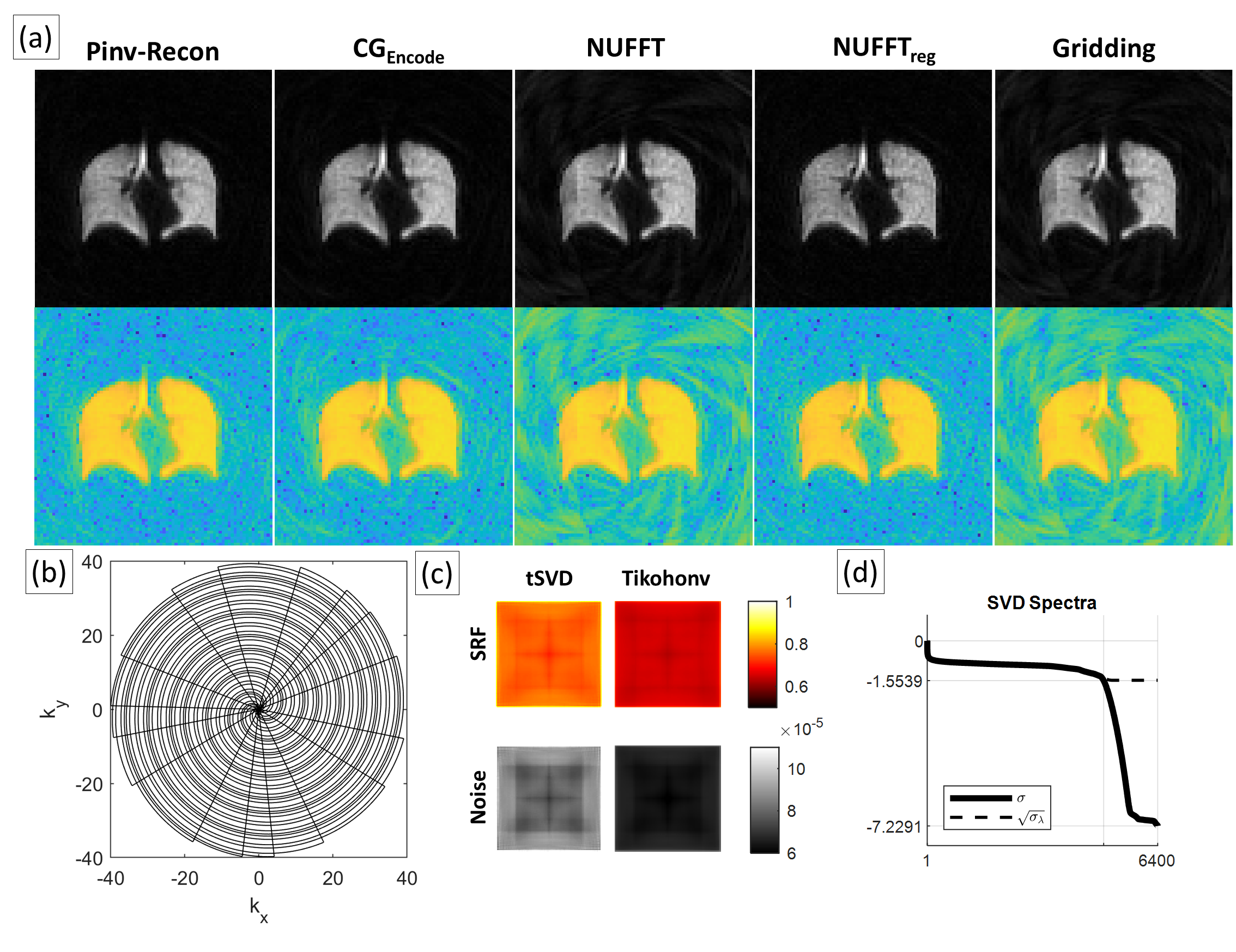}
    \caption{(a) Reconstruction of stack-of-spirals data using, from left to right, Pinv-Recon, $CG_{Encode}$, NUFFT, NUFFT with regularization, and gridding. Gridding and NUFFT results in streaking artifacts that can be reduced with regularization, whilst Pinv-Recon eliminates these artifacts and boosts SNR. First row shows the images in the original scale, and second row shows them in log scale. (b) The golden angle spiral in  one Z phase-encode of the stack-of-spirals trajectory (c) Top row: the SRFs for tSVD regularization and Tikhonov regularization respectively. Bottom row: the corresponding noise matrices (d) The singular values of \textbf{Encode}, with the singular values of the Gram matrix after Tikhonov regularization shown by the dotted line.}
    \label{fig:xenon}
\end{figure}

As shown in Fig. \ref{fig:xenon}, a hyperpolarized Xenon-129 dataset (acquired using a golden-angle stack-of-spirals with MTX 80 $\times$ 80 $\times$ 41) demonstrates Pinv-Recon at a larger matrix size and with a separable Z-dimension. This is an example of non-Cartesian in-plane sampling combined with Cartesian phase-encoding along Z with a single slab-selection pulse. In reconstruction, the uniformly sampled Z-dimension was processed separately: Pinv-Recon was applied only along Z without regularization (equivalent to a standard 1D FFT), while a separate Pinv-Recon was applied to the non-uniformly sampled in-plane 2D spiral. Reconstruction at the native resolution and over-discretized to twice the resolution can be performed at reasonable compute times of just a couple of seconds using Cholesky decomposition.

Pinv-Recon was also compared against several methods: (1) $CG_{Encode}$, (2) non-uniform FFT (NUFFT) reconstruction \cite{fesslerMichiganImageReconstruction} (3) NUFFT reconstruction with additional regularization, and (4) conventional gridding. The benefit of inherent regularization in Pinv-Recon is highlighted, allowing better reconstruction of ill-conditioned undersampled data \cite{ueckerRealtimeMRIResolution2010}. The golden-angle spiral trajectory had unevenly-spaced  interleaves, creating voids in k-space that translate to density compensation factors larger than unity. This creates an ill-posed problem which gridding algorithms may not be able to solve properly without regularization. Streaking artifacts that were present in the gridding and the NUFFT reconstructed images can be minimized with additional regularization. However, this is inherently accounted for in the Pinv-Recon and $CG_{Encode}$ implementations with Tikhonov regularization included.

The SRF (Fig. \ref{fig:xenon}c, top row) and the noise matrix (Fig. \ref{fig:xenon}c, bottom row) for the reconstruction of the hyperpolarized Xenon image can be calculated for tSVD and Tikhonov regularization respectively, showing that the image is less well-represented and less noisy towards the center of the image with this level of Tikhonov regularization.

\subsection{High Resolution 2D Cartesian VDPD ($T_2$-weighted Knee)}
\begin{figure}[h!]    \centering
    \includegraphics[width=\linewidth]{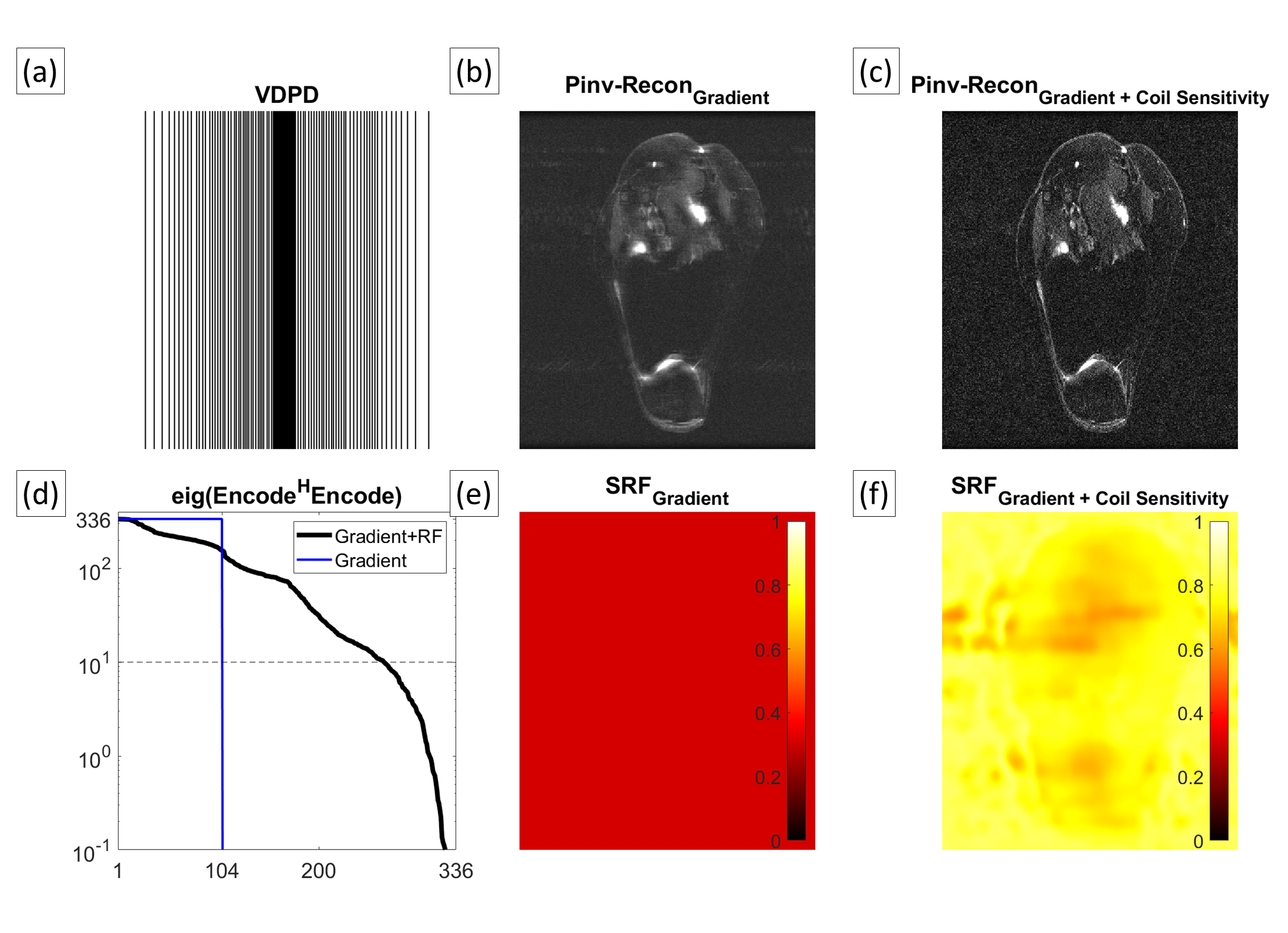}
    \caption{High Resolution $T_2$-weighted knee image reconstructed with Pinv-Recon. (a) Plot of the k-space trajectory, showing variable sampling along the phase encoding dimension from left to right (b) Reconstructed image using gradient encoding only (c) Reconstructed image using gradient and coil sensitivity encoding (d) Eigenvalue spectrum, for gradient encoding only (blue) and for gradient encoding with coil sensitivity encoding (black) (e) SRF for reconstruction with gradient encoding only (f) SRF for reconstruction with gradient encoding and coil sensitivity encoding.}
    \label{fig:knee}
\end{figure}

We further demonstrate the efficiency of Pinv-Recon for separable imaging dimensions, using a high resolution 18-channel 2D Variable Density Poisson Disk (VDPD) knee dataset (MTX 384 $\times$ 336), which is evenly sampled in the readout dimension, but variably sampled in the phase encoding dimension (Fig. \ref{fig:knee}). The dataset undersamples k-space in the phase encoding dimension with an acceleration factor (R) of around 3. Along the frequency encoding line, each phase encode is reconstructed separately, accounting for coil sensitivity. 

We show that by incorporating the coil sensitivity profiles, a higher resolution and more accurate image can be reconstructed by ameliorating pseudo-random sub-sampling artifacts (Fig. \ref{fig:knee}c). From the corresponding SRF map with values close to 1 (Fig. \ref{fig:knee}f), it can be seen that almost full nominal image resolution can be recovered, in contrast to the gradient-encoding-only SRF map with values of around 0.3 (Fig. \ref{fig:knee}e).

\subsection{High Resolution 2D Radial with Undersampling (CG-SENSE Challenge Dataset)}
We explored the application of Pinv-Recon to even larger matrix sizes, demonstrating its computational feasibility using the widely recognized open-source radial brain dataset from the ISMRM CG-SENSE reproducibility challenge \cite{maierCGSENSERevisitedResults2021}. Using this dataset, we show that a MTX 320 $\times$ 320 image can be reconstructed in just a few minutes, with the possibility of including non-Cartesian SENSE reconstruction for undersampled data. The size of the encoding matrix can be reduced by masking out image voxels beyond the field-of-view of interest. A circular mask with diameter equal to the side of the image was used initially, providing an image to which a signal threshold can be applied to generate an even smaller mask for subsequent reconstructions. This approach reduces the number of unknowns without discarding image content.

\begin{figure}[h!]    \centering
    \includegraphics[width=\linewidth]{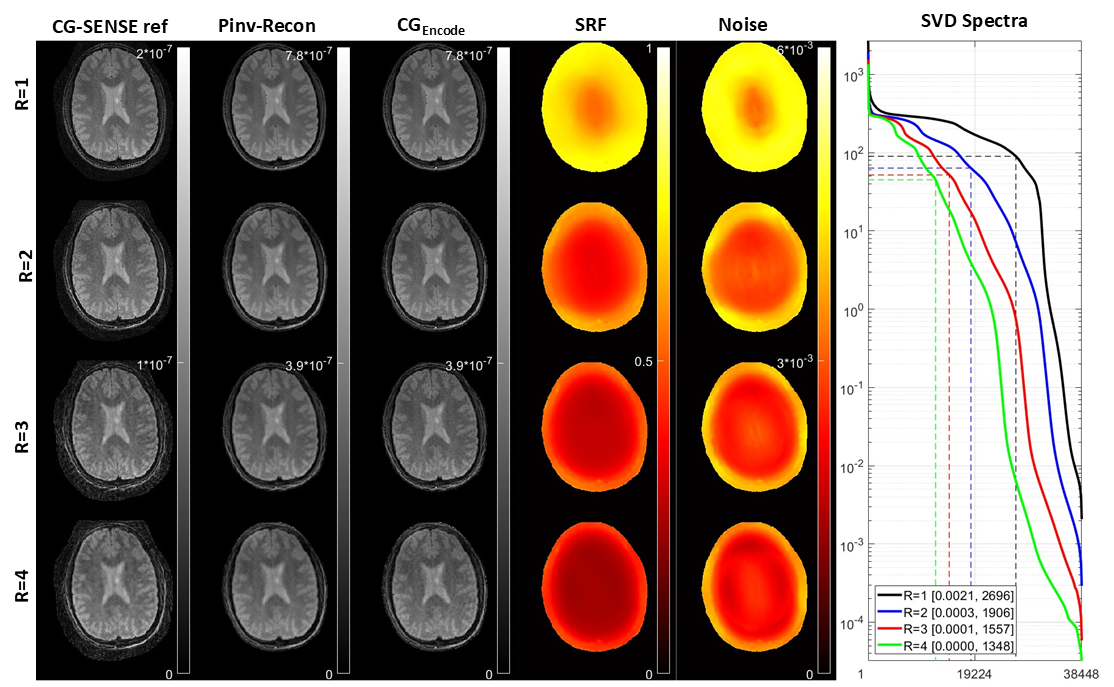}
    \caption{CG-SENSE reproducibility challenge dataset. From top to bottom: undersampling factors from 1 to 4. From left to right: reconstruction using CG-SENSE reference implementation in MATLAB, reconstruction using Pinv-Recon with coil sensitivity encoding, reconstruction using $CG_{Encode}$, SRF maps, noise map, SVD spectra.}
    \label{fig:cgsense}
\end{figure}

Instead of using forward and inverse gridding or NUFFT to iteratively reconstruct the image, we demonstrate the computational feasibility of Pinv-Recon as an alternative to non-Cartesian SENSE by simply incorporating the coil sensitivity into the encoding matrix and pseudoinverting it. Reconstruction with acceleration factors of R = 1 to 4 are shown in Fig. \ref{fig:cgsense}. $CG_{Encode}$ gives a similar image to Pinv-Recon but does not straightforwardly provide the SRFs and noise maps (columns 4 and 5 in Fig. \ref{fig:cgsense}). Especially in the context of non-Cartesian undersampled reconstruction, these metrics provide valuable information about the optimality of encoding and reconstruction. Using the SVD, whilst computationally more expensive than other matrix decomposition methods, can still be useful for shedding light onto the conditioning of the problem, as shown in the rightmost panel of Fig. \ref{fig:cgsense}.

\subsection{High Resolution 3D Stack-of-Stars (Abdominal Imaging)}

\begin{figure}[h!]    \centering
    \includegraphics[width=\linewidth]{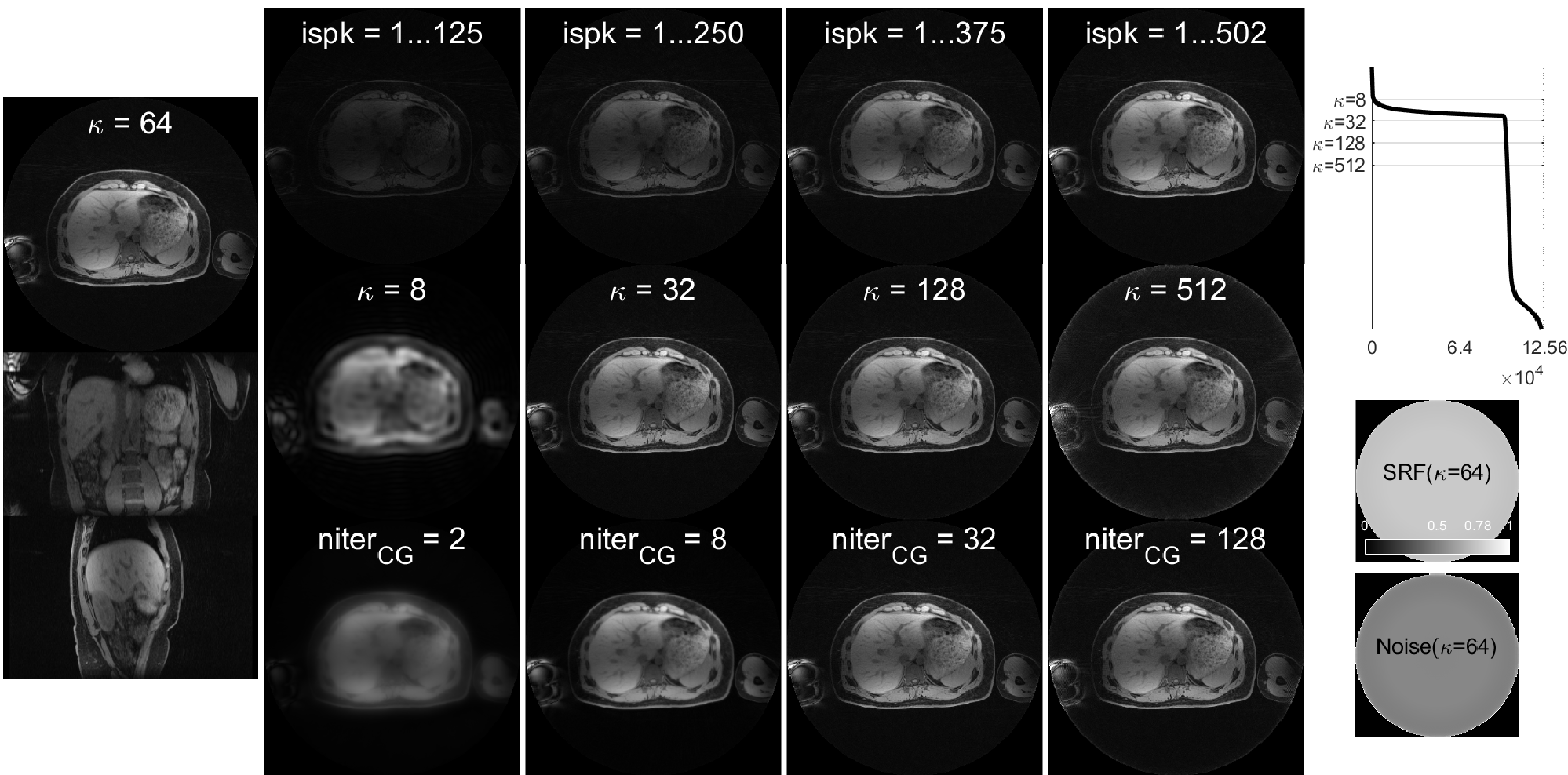}
    \caption{Left: Axial, coronal and saggital views of the 3D stack-of-stars abdominal imaging dataset. In the middle four columns, top: varying the number of spokes used in reconstruction, middle: varying the condition number $\kappa$ by changing the tSVD threshold, bottom: varying the number of iterations in $CG_{Encode}$. Right: SVD spectrum, SRF map, and noise map.}
    \label{fig:lavastar}
\end{figure}

Separable 3D Pinv-Recon was demonstrated using a high-resolution abdominal 3D stack-of-stars spoiled gradient echo acquisition (Fig. \ref{fig:lavastar}) which has MTX 320 $\times$ 320 $\times$ 144. Similar to the ISMRM CG-SENSE reproducibility challenge, only image pixels in the masked region are included in reconstruction. The Z phase-encoding dimension was factored out via a 1D FFT, and the XY dimensions were reconstructed using Pinv-Recon, forming a 3D image which is shown in the leftmost column in axial, coronal and sagittal views.

The potential for real-time image reconstruction, i.e., updating the reconstructed image during scanning, is demonstrated in this dataset. Unlike gridding and iterative reconstruction methods, Pinv-Recon is data agnostic in that the reconstruction matrix can be calculated simply using the known trajectory, even before data is acquired. Once data acquisition finishes, only a simple matrix-vector reconstruction remains. The top row of Fig. \ref{fig:lavastar} shows images obtained using increasing fractions of the available data spokes from left to right, indicating the potential for real-time updates during imaging. The acquisition time of certain MRI scans can last up to several minutes (6 minutes for this dataset), and real-time updates can be useful for radiographers or scanner operators to monitor the acquisition.

Comparison of Pinv-Recon and $CG_{Encode}$ is shown in the bottom two rows of Fig. \ref{fig:lavastar}. The middle row shows images obtained using varying thresholds for Pinv-Recon with tSVD, demonstrating the trade-off between resolution and SNR offered by varying the condition number $\kappa$, also illustrated in the SVD spectrum on the right. The bottom row shows images reconstructed using $CG_{Encode}$. As is common with CG-based iterative methods, image quality first improves, peaks and then degrades with an increasing number of iterations.

\section{Discussion}

With the drastic increase in computational power over the past two decades \cite{mooreCrammingMoreComponents1965}, inverting large encoding matrices has become much more feasible. For example, Van de Walle \textit{et al.} (2000) reported approximately 75 minutes to reconstruct a MTX 64 $\times$ 64 image ($N_k \times N_r$ = 64$^2$ $\times$ 4033) \cite{vandewalleReconstructionMRImages2000}; today, the same size matrix can be decomposed in seconds (approximately 2~s with SVD or less than 0.1~s with Cholesky on our workstation). In our experiments, a $\sim$2~GB encoding matrix (roughly equivalent to a MTX 128 $\times$ 128 image with full Fourier encoding) was inverted via Cholesky Decomposition in about 0.4~s. Memory limitations, a major barrier in the past, are far less restrictive on modern 64-bit systems that can address terabytes of RAM \cite{wangComputerMemoryStorage2021,mathworksTableMatlabRelease2024}. Looking forward, ongoing developments in high-performance computing architectures, for example larger GPU memory \cite{NVIDIABlackwellArchitecture}, extreme parallelism with AI chips \cite{ProductChipURL} and half-precision or mixed-precision arithmetic\cite{mathworksWhatHalfPrecision}, promise to further extend the capabilities of Pinv-Recon. Mixed-precision computations, in particular, offers an interesting approach to speed-up computations by using lower precision for initial calculations but storing results at higher precision, increasing computational efficiency whilst limiting the compromise on accuracy \cite{highamMixedPrecisionAlgorithms2022}. Moreover, modern randomized algorithms for low-rank matrix approximation, such as the \texttt{svdsketch} function in MATLAB\cite{mathworksSvdsketchComputeSVD}, are emerging as potential ways to approximate the pseudoinverse more quickly by sketching the dominant subspace. While such approaches were not used in this work, they represent promising avenues for handling extremely large encoding matrices by trading off some accuracy for speed.

Table 1 summarizes the various algebraic methods and computational strategies mentioned, their applicable scenarios, as well as their advantages and limitations. Among the matrix inversion methods examined, SVD is the most robust and provides valuable insight (via the SVD spectrum) into the conditioning of the problem \cite{vandewalleReconstructionMRImages2000,desplanquesIterativeReconstructionMagnetic2002}. However, SVD is computationally expensive. Our exploration of alternative factorizations revealed that many of these methods, though not previously considered in MRI reconstruction, can offer substantial speed-ups. In particular, Cholesky decomposition was observed to be over 100$\times$ faster than SVD in some cases. These methods operate on the normal equations (decomposing the Gram matrix), effectively reducing the problem size to the smaller dimension of \textbf{Encode}. Depending on the problem size relative to the available compute resources, different computational strategies may be considered. For very small encoding matrices, CPU computations are preferred. GPU acceleration is useful for larger encoding matrices, and block-wise inversion on the GPU is preferred for even larger encoding matrices that do not fit on the GPU memory. Other strategies to better manage the inversion include masking voxels outside region-of-interest, computing in single precision, reducing the problem size by separating uniformly sampled dimensions, and procedural construction of the Gram matrix with coil sensitivity encoding.

Perhaps the greatest strength of Pinv-Recon is its simplicity and versatility. It seamlessly accommodates non-Cartesian trajectories and a variety of corrections that would otherwise require complex dedicated algorithms \cite{wrightNonCartesianParallelImaging2014,zhuHybridSpaceSENSEReconstruction2016,harkinsIterativeMethodPredistortion2014,ueckerESPIRiTEigenvalueApproach2014}. Conventional reconstruction pipelines often involve multiple stages (e.g., density compensation \cite{rascheResamplingDataArbitrary1999,tanPointSpreadFunction2005}, trajectory correction, off-resonance modulation, motion gating, coil combination, deapodization, gradient nonlinearity correction), each adding complexity and variability. Many of these algorithms cannot be easily combined with one another. In Pinv-Recon, all these effects are incorporated into the single encoding matrix, thereby simplifying the reconstruction workflow and improving reproducibility. Iterative methods can, in principle, also incorporate such factors, as exemplified by previous work using the explicit encoding matrix in conjugate-gradient methods, for example Kadah and Hu's \textit{Algebraic Reconstruction for Magnetic Resonance Imaging Under $B_0$ Inhomogeneity}\cite{kadahAlgebraicReconstructionMagnetic1998} and Wilm \textit{et al.}'s \textit{Higher order reconstruction for MRI in the presence of spatiotemporal field perturbations} \cite{wilmHigherOrderReconstruction2011}. Leveraging modern compute resources, Pinv-Recon implements this more simply, and also calculates \textbf{Recon} to provide, by simple matrix-matrix multiplication, additional metrics such as SRF maps and noise amplification maps that can serve as quality assurance measures and guide future sequence design \cite{pruessmannSENSESensitivityEncoding1999,sanchez-gonzalezMinimumnormReconstructionSensitivityencoded2006,roemerNMRPhasedArray1990,kellmanImageReconstructionSNR2005,tunnicliffeUseNoiseCovariance2011}. Unlike iterative reconstructions, which require full data before processing, Pinv-Recon can pre-compute \textbf{Recon} and reuse it across acquisitions with the same encoding, offering the option of real-time updates during acquisition and leaving only a matrix-vector multiplication after acquisition. Regularization---implemented transparently via tSVD, Tikhonov regularization, or both---further reduces noise and stabilizes ill-conditioned, undersampled problems, as demonstrated in our Xenon-129 dataset. Pinv-Recon can be further extended to non-linear reconstructions\cite{ueckerImageReconstructionRegularized2008}, for example compressed-sensing problems, where k-space is randomly subsampled \cite{lustigSparseMRIApplication2007}, as well as reconstruction using implicit neural representation \cite{zhuImplicitNeuralRepresentation2025}.

Recent work related to pseudoinversion in MRI reconstruction has also investigated various matrix decomposition techniques to facilitate efficient inversion, including QR decomposition and Jacobi SVD \cite{ullahQRdecompositionBasedSENSE2018,qaziSingularValueDecomposition2017}. These methods typically focus on unaliasing undersampled Cartesian images by inverting much smaller coil sensitivity encoding matrices after an initial reconstruction using gradient encoding. In contrast, the present work directly inverts the full encoding matrix (including gradient and coil sensitivity terms) and extends this framework to non-Cartesian sampling schemes.

With the move towards open-source and reproducible MR acquisition and reconstruction, as evidenced by the growing popularity of the vendor-neutral pulse sequence programming environment Pulseq, Pinv-Recon shows great promise as a simple, versatile, and reliable reconstruction method \cite{laytonPulseqRapidHardwareindependent2017,veldmannOpensourceMRImaging2022}. An example implementation of Pinv-Recon for Pulseq is now available and can be found online \cite{SequenceExamplesGEPge2Spiral}. Researchers developing novel reconstruction techniques may use Pinv-Recon as a reference reconstruction method since it provides the generic linear-least squares solution (or the minimum 2-norm solution for underdetermined cases). Whilst the simulations and characterization in this manuscript were run in MATLAB for its efficient matrix operations, we have also included Python-equivalent code in the GitHub repository linked in the Data Availability Statement, to provide an option to run the code in  an open-source programming language. Moreover, Pinv-Recon can also be used as an instructive educational tool for students familiarizing themselves with various aspects of MR physics, exploiting its modularity to visualize each encoding mechanism. 

We acknowledge that Pinv-Recon is not intended to replace all FFT-based or iterative methods; rather, it excels in small to medium matrix size applications (up to approximately 256 $\times$ 256 in-plane) where its computational burden is manageable and its advantages in flexibility and reproducibility are most pronounced. For very high-resolution imaging, iterative methods may remain more practical.

In conclusion, we have shown that direct pseudoinversion of the encoding matrix---once considered impractical---is now within reach for many MR reconstruction problems. Pinv-Recon leverages modern computational resources and optimized linear algebra routines to provide a simple, linear, and extendable reconstruction framework that is particularly well-suited for MRI applications with small to medium matrix sizes and specialized encoding requirements. The presented framework can unify multiple linear encoding strategies and correction mechanisms into a single, comprehensive algebraic model. This one-step approach not only computes the image directly but also yields valuable image metrics such as spatial response functions and noise maps, and has the potential for real-time feedback during imaging, making it a valuable tool for both research and clinical applications.

\section{Methods}
\subsection{Computational Evaluation and Numerical Simulations}
The computational evaluation and numerical simulations were performed as described in Section 3, using Matlab 2023b (The MathWorks, MA, USA). The MATLAB implementation can be also found in the Git repository linked in the Data Availability Statement.

\subsection{Applications}
All reconstructions were performed using the equations outlined in Section 2. Details of the acquisition of each dataset are as follows:

\textbf{Hyperpolarized Carbon-13 Dataset:} A healthy volunteer was scanned after IV injection of 35~mL ofhyperpolarized [1-\(^{13}\)C]pyruvate (polarized for approximately 4 hours in the SPINlab (GE Healthcare, WI, USA)). The acquisition was a spectral-spatial single-arm spiral sequence covering a 240 × 240~mm field-of-view with eight 20~mm slices; flip angles were 5° for pyruvate, 15° for lactate, and 60° for bicarbonate, acquired at 4~s intervals. The scan was performed on a GE 3T Premier (GE Healthcare, WI, USA)  system using an 8-channel transmit-receive \(^{13}\)C/\(^{1}\)H head coil (Rapid Biomedical, Rimpar, Germany). Coil sensitivity maps were estimated via polynomial smoothing of pyruvate images from each individual coil and normalized by the sum-of-squares of all coils. This project was approved by the University of Nottingham Medical School Ethics Committee (Ref: 416-1911 BRAIN DNP) and was conducted in accordance with the standards set by the
latest version of the Declaration of Helsinki, except for registration in a database. Informed, written consent was obtained from the volunteer before participation.

\textbf{Hyperoplarized Xenon-129 Dataset:} A healthy 34-year-old female participant inhaled 1~L of enriched \(^{129}\)Xe gas (polarized for approximately 30 minutes). Imaging was performed on a 3T GE Premier scanner with a flexible \(^{129}\)Xe chest coil (PulseTeq, Cobham, UK) using a stack-of-spirals 3D sequence covering a 400 × 400 × 200~mm field-of-view. The sequence employed 16 spiral interleaves with golden-angle rotation, a bandwidth of 250~kHz, TR = 15.6~ms, TE = 2.3~ms, and a hard excitation pulse yielding an effective flip angle of approximately 66° per volume. Participants gave informed written consent, and ethics approval was provided by the University of Oxford
Central University Research Ethics Approval (Reference: R77150/RE001).

Pinv-Recon of the dataset was compared with conjugate gradient iterative reconstruction with 11 iterations, NUFFT using the Michigan Image Reconstruction Toolbox (MIRT) \cite{fesslerMichiganImageReconstruction}(density compensation calculated using the Pipe and Menon method \cite{pipeSamplingDensityCompensation1999}, NUFFT object calculated with minmax interpolator with Kaiser-Bessel scaling, kernel width of 6, table-based interpolation with a $2^{10}$ oversampling factor) without and with additional regularization with quadratic penalized weighted least squares (10 iterations, with a beta of $2^{-21}$), and conventional gridding (Kaiser-Bessel filtering \cite{beattyRapidGriddingReconstruction2005} with kernel width of 3 and oversampling to 1.125 of the original grid size, density compensation factor determined using Voronoi triangulation).

\textbf{$T_2$-weighted Knee:} The knee dataset was acquired with a dedicated knee coil with 18 receive channels using a 2D fast recovery fast spin echo (frFSE) acquisition with fat saturation. The acquisition parameters were TR = 6 s, TE = 108 ms. The trajectory was a 1D variable density poison disc (VDPD) sampling with image matrix size of 384 in the readout direction and 336 in the phase encoding direction. The phase encoding dimension was acquired with 3.23 acceleration (104/336), and the center of k-space (-12:12) was fully sampled for coil sensitivity calibration.

\textbf{ISMRM CG-SENSE Challenge Dataset:} The ISMRM CG-SENSE Challenge dataset was obtained from \textit{CG-SENSE revisited: Results from the first ISMRM reproducibility challenge} \cite{maierCGSENSERevisitedResults2021}.

\textbf{Stack-of-stars Abdominal dataset:} For the 3D stack-of-stars dataset (using the Liver Acquisition with Volume Acceleration LAVA-Star sequence), the acquisition parameters were FOV = 40.0 $\times$ 40.0 $\times$ 31.68 cm, resolution = 1.25 $\times$ 1.25 $\times$ 2.2 mm, TE/TR = 1.488/3.276 ms, FA = 12$^{\circ}$, BW = $\pm$62.5 kHz, with interleaved fat suppression, resulting in a scan time of 5 min 58 sec.

\bibliographystyle{unsrt}
\bibliography{PINV_scientific_reports}

\section*{Funding Declaration}
K.Y. acknowledges an Oxford-Medical Research Council Doctoral Training Partnership iCASE award, the Oxford-Radcliffe Scholarship, and GE HealthCare for graduate funding. D.J.T. was funded by a British Heart Foundation Senior Basic Science Research Fellowship (FS/19/18/34252). J.T.G. is funded by the Oxford Biomedical Research Centre, and acknowledges the BHF Centre of Research Excellence, University of Oxford for funding. The hyperpolarized Xenon-129 datasets were funded by the British Heart Foundation (RE/18/3/34214) and the hyperpolarized Carbon-13 dataset was funded by Nottingham Life cycle 5.

\section*{Author Contributions Statement}
K.Y.: writing - original draft (lead); conceptualization (supporting); data collection (equal); methodology (equal); software (equal); formal analysis (equal); writing - review and editing (equal). C.T.: conceptualization (supporting); writing - review and editing (equal). R.F.S.: conceptualization (supporting); writing - review and editing (equal). B.W.: visualization; writing - review and editing (equal). A.M.: data collection (equal). S.S.: data collection (equal); writing - review and editing (equal). P.M.: data collection (equal). D.A.: data collection (equal). F.V.G.: funding acquisition (equal); supervision. D.J.T.: funding acquisition (equal); supervision (equal); writing - review and editing (equal). J.T.G.: funding acquisition (equal); supervision (equal); writing - review and editing (equal). F.W.: conceptualization (lead); writing - original draft (supporting); methodology (equal); software (equal); formal analysis (equal); supervision (equal); writing - review and editing (equal). All authors read and approved the final manuscript.

\section*{Competing Interests Statement}
C.T. is an employee of The Mathworks. R.F.S. and F.W. are employees of GE HealthCare. S.S. is an Editorial Board Member of \textit{Scientific Reports}. The remaining authors declare no relevant conflicts of interest.

\section*{Data Availability Statement}
The code supporting the numerical simulations in this manuscript, as well as some example datasets can be found at https://github.com/univ39/Pinv-Recon.

\flushbottom
\newpage

\flushbottom
\newpage

\flushbottom
\newpage

\flushbottom
\newpage

\flushbottom
\newpage

\flushbottom
\newpage

\flushbottom
\newpage

\flushbottom
\newpage

\flushbottom
\newpage
\begin{figure}[h!]    \centering
    \includegraphics[width=0.95\linewidth]{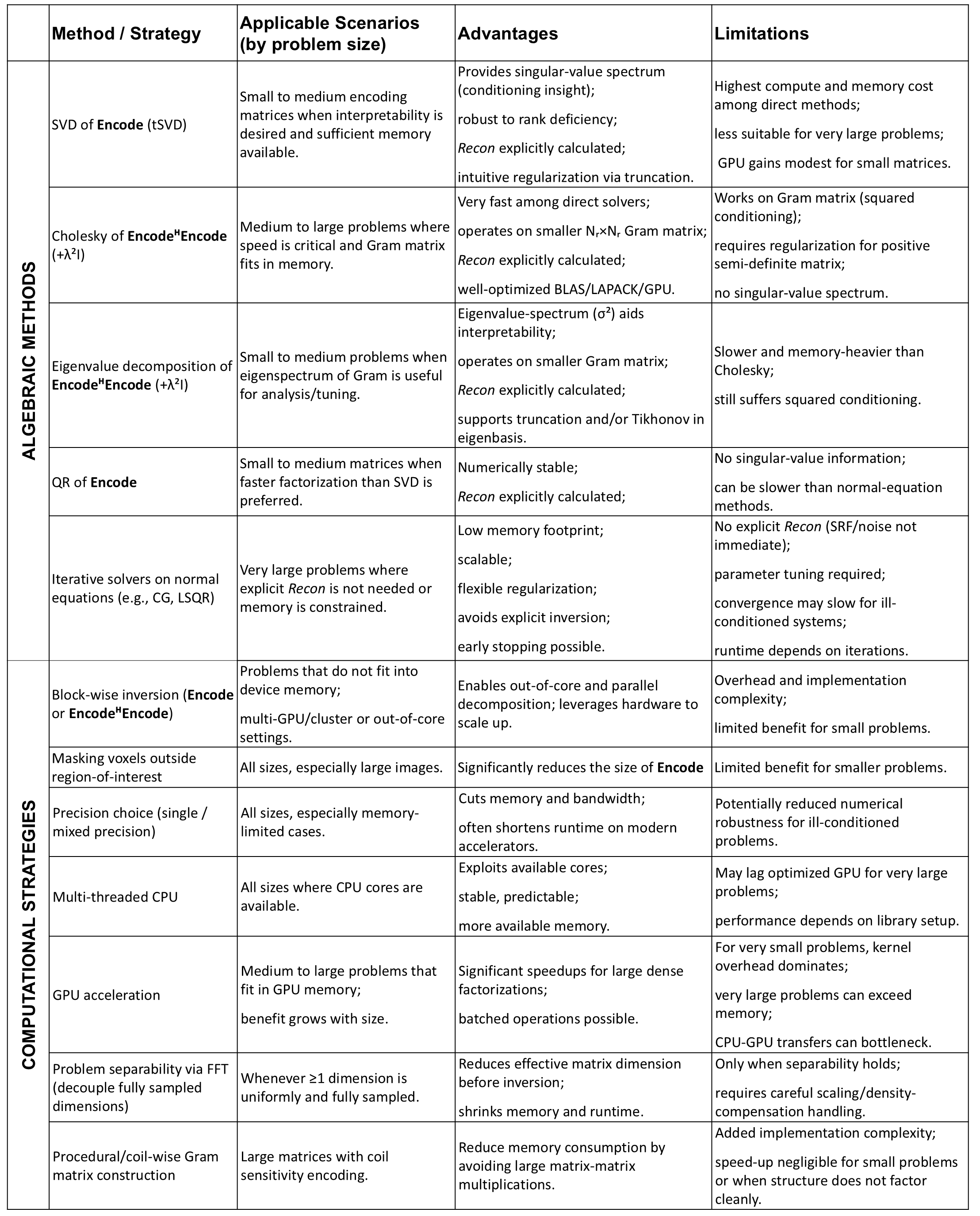}
    \small{\textbf{Table 1.} Table summarizing algebraic methods and computational strategies, their applicable scenarios, advantages and limitations respectively. Notes: (i) “Small/medium/large” refer to the size of the encoding problem relative to available CPU/GPU memory and bandwidth. (ii) Methods that explicitly compute the reconstruction matrix ($\textbf{Recon} = \textbf{Encode}^{+}$) enable direct calculation of SRF and noise matrices, the reusing of \textbf{Recon}, and real-time feedback during acquisition. (iii) Normal-equation / Gram-matrix methods square the singular values (eigenvalues of $\textbf{Encode}^H\textbf{Encode}$), worsening conditioning; Tikhonov regularization ($\lambda^2I$) is recommended to ensure positive-definiteness and improve stability.}
    \label{fig:table}
\end{figure}

\setcounter{figure}{0} 
\setcounter{section}{0} 
\renewcommand{\thefigure}{S\arabic{figure}}

\clearpage

\section*{Supplementary Materials}

\section{Versatility of Pinv-Recon for Different K-Space Trajectories}
\begin{figure}[H]
    \centering
    \includegraphics[width=0.85\linewidth]{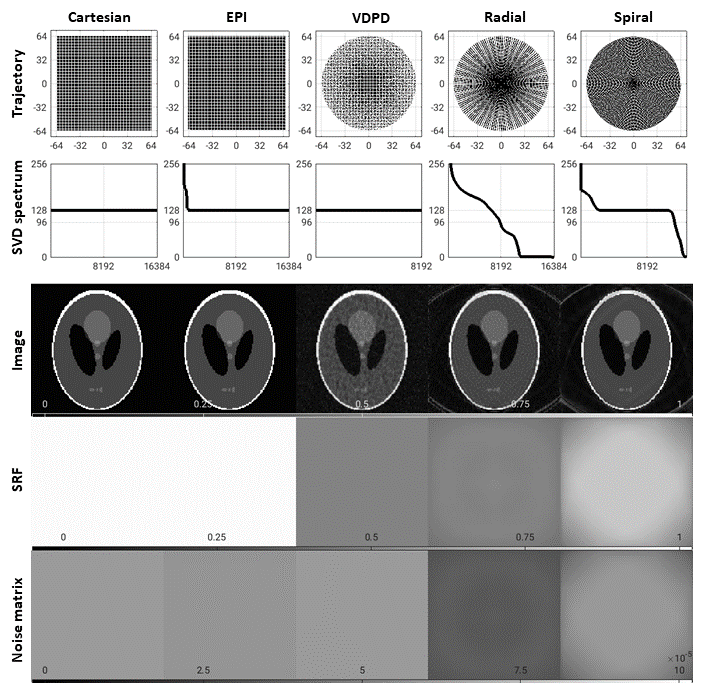}
    \caption{Demonstration of Pinv-Recon on a variety of k-space sampling trajectories. From left to right: Cartesian, EPI, Variable-Density Poisson Disk (VDPD), radial, spiral. From top to bottom: sampling trajectory, SVD, reconstructed Shepp-Logan image, SRF, noise matrix.}
    \label{fig:supp_sl}
\end{figure}

Figure~\ref{fig:supp_sl} demonstrates the versatility of Pinv-Recon for different k-space sampling trajectories, plotting from top to bottom the trajectory, its SVD spectrum, the reconstructed image, the SRF maps, and the Noise matrices.
The SVD spectra show that Cartesian sampling is the best-conditioned, as seen also in its artifact-free reconstruction of the Shepp-Logan phantom. Pinv-Recon also allowed the direct calculation of the SRFs of each trajectory, showing all pixels are fully represented in the reconstructed image. The noise matrices reflect that noise is amplified in the edges of the reconstructed image for the radial and spiral trajectories.

\newpage

\section{Computational Evaluation on a Mobile Workstation}
Section 3.1 listed the computational times required for a high-performance workstation similar to the reconstruction machines found on MRI scanners. Here, we show that even modern mobile workstations with lower-performance can manage the direct pseudoinversion of various encoding matrices.
We perform the same computational evaluation described in section 3.1.1 on a Precision 7680  mobile workstation(Dell, Texas, USA), which has 13th Gen Intel\textregistered Core$^{\mathrm{TM}}$ i9-13950HX, 2200 Mhz, 24 Core(s), 32 Logical Processor(s), 64.0 GB of Installed Physical Memory (RAM), and an NVIDIA\textregistered RTX$^{\mathrm{TM}}$ 2000 Ada Generation Laptop GPU with 16GB.

\begin{figure}[H]
    \centering
    \includegraphics[width=0.9\linewidth]{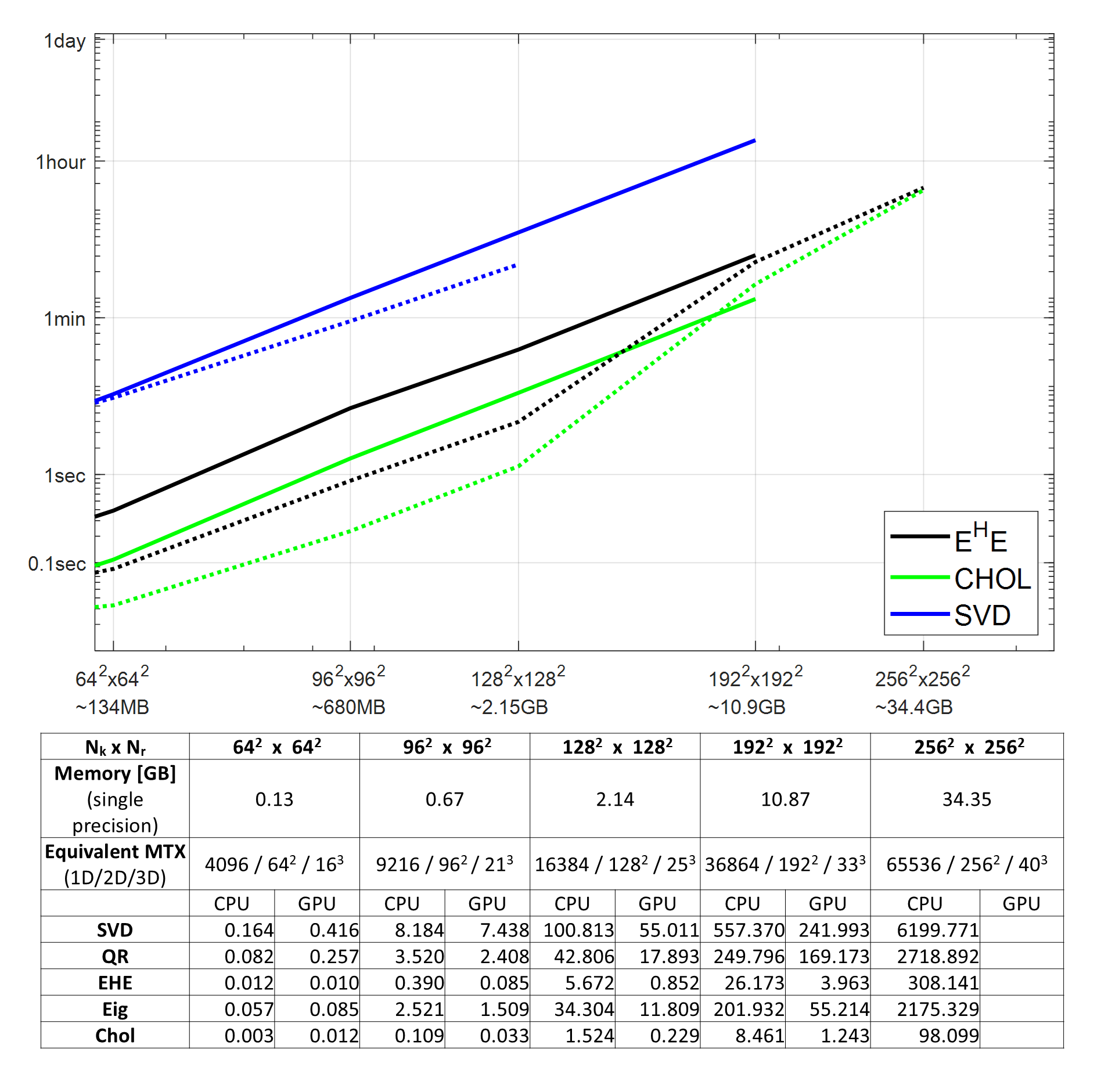}
    \caption{Graph showing computational times for matrix decomposition on a mobile workstation}
    \label{fig:supp_mobile}
\end{figure}

\section{$B_0$ Correction Examples}
\begin{figure}[H]
    \centering
    \includegraphics[width=0.8\linewidth]{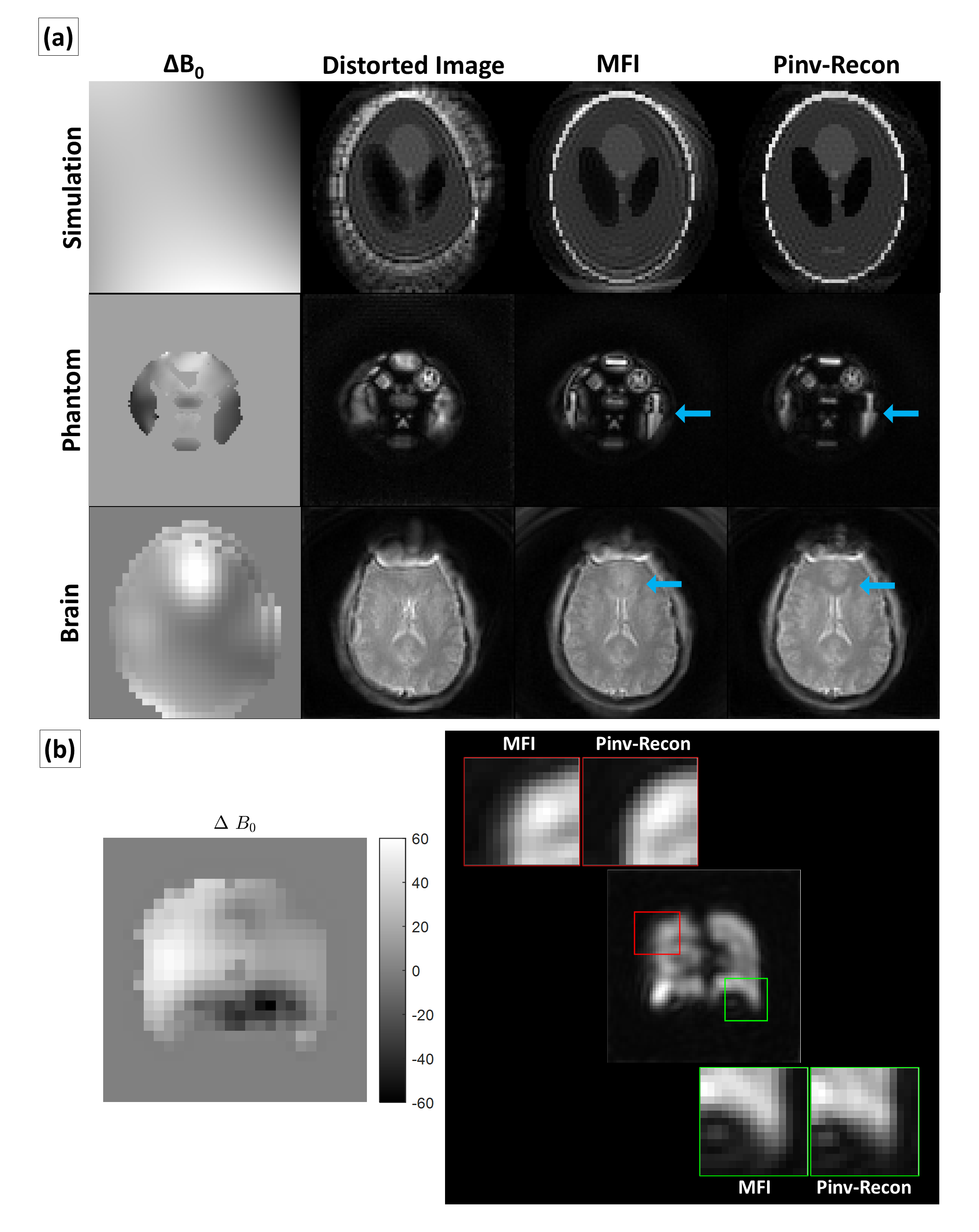}
    \caption{(a) Shepp-Logan simulations, Structural Phantom acquisitions, and in vivo brain acquisitions comparing correction of $B_0$ distortion using Pinv-Recon versus MFI  (b) Left: Off-Resonance Map. Right: Low-resolution hyperpolarized Xenon-129 image, with the original image shown in the center and the $B_0$ corrected images using MFI and Pinv-Recon shown outside}
    \label{fig:supp_b0}
\end{figure}
\begin{table}[H]
    \centering
    \includegraphics[width=\linewidth]{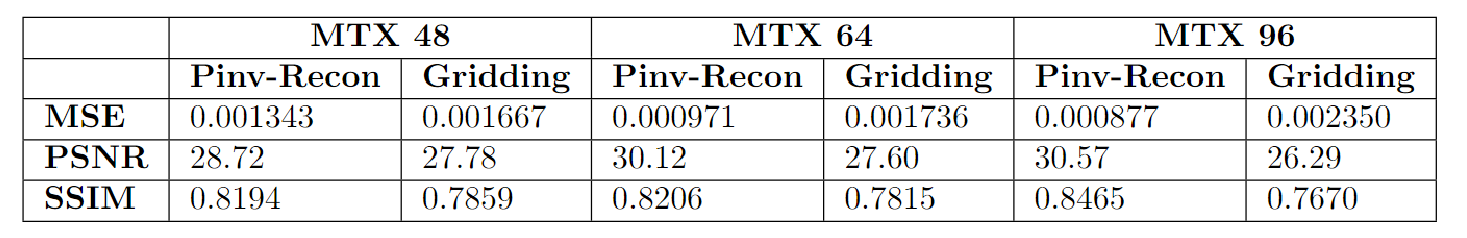}
\caption{Results comparing $B_0$ correction using Pinv-Recon versus using gridding with MFI correction in a structured resolution phantom.}
\label{table:phantom_B0}
\end{table}

Figure~\ref{fig:supp_b0} shows two examples of including $B_0$ correction in the Pinv-Recon encoding matrix, compared to $B_0$ with Multifrequency Interpolation (MFI) \cite{manMultifrequencyInterpolationFast1997}. Figure~\ref{fig:supp_b0}a shows $B_0$ correction in a Shepp-Logan simulation, a structured resolution phantom and in vivo a human brain scan at the proton frequency. The off-resonance maps are shown in column 1, the distorted images are shown in column 2. By incorporating the $B_0$ maps into Pinv-Recon and into an MFI correction for the image obtained through gridding reconstruction, the blurring effects can be ameliorated, recovering images close to the reference image. Pinv-Recon results in lower MSE, higher PSNR, and higher SSIM than MFI for all of the matrix sizes (Table~\ref{table:phantom_B0}). 

Figure~\ref{fig:supp_b0}b shows low resolution 2D spiral hyperpolarized Xenon-129 lung dataset, which had a longer readout time. Off-resonance blurring can be observed in the original image. Using gridding reconstruction with MFI correction and using Pinv-Recon with embedded $B_0$ encoding improved image quality, but Pinv-Recon sharpens the image to a greater extent. 

The acquisition details for the proton dataset were maximum gradient amplitude = 30 mT/m; slew rate = 120 T/m/s; FOV = 240$\times$240 mm; bandwidth = 250 kHz. Acquisition parameters: Flip Angle = 30$^\circ$; TR = minimum TR; Slice thickness = 3mm). The matrix 48 four-arm spiral was repeated at five different TEs ([0,1,2,5,10]ms) to iteratively fit for a $B_0$ map using the MEDI toolbox. The phantom was imaged in a 3T GE Premier scanner and using a 5-channel flexible AIR coil (GE HealthCare, WI). The healthy volunteer (Male; Age 30) was scanned in the same scanner using a 16-channel receive-only headcoil (GE HealthCare, WI), using the same single-arm MTX96 spiral and the four-arm MTX 48 spiral.

Those for the low-resolution 2D spiral integrated into the Transmit Gain Calibration of a hyperpolarized Xenon-129 scan, which used a 10\% dose of Xenon. The participant (Female; Age 24) inhaled 1 L hyperpolarized gas containing a mixture of xenon, polarized for $\sim$10 min, and nitrogen (0.1:0.9 L, respectively). FOV = 400$\times$400 mm, 1.8 ms partially self-refocused excitation pulse, TR = 230 ms, bandwidth = 250 kHz. The $B_0$ map was determined by repeating this acquisition at echo times of 1.4, 2.3, 3.9, 6.5, 10.8, 18, 30, 50ms then fitting with the MEDI toolbox.
\newpage
\section{Arterial Spin Labelling Example}
Other suitable applications for Pinv-Recon are medium-resolution functional imaging applications such as Arterial Spin Labelling (ASL). This figure shows an example of reconstructing a  stack-of-spirals ASL dataset (MTX 128$^2$ $\times$ 42, FOV = 240 $\times$ 240 mm, 8 arms, TR = 4.8 s, TE = 10 ms).

\begin{figure}[H]
    \centering
    \includegraphics[width=\linewidth]{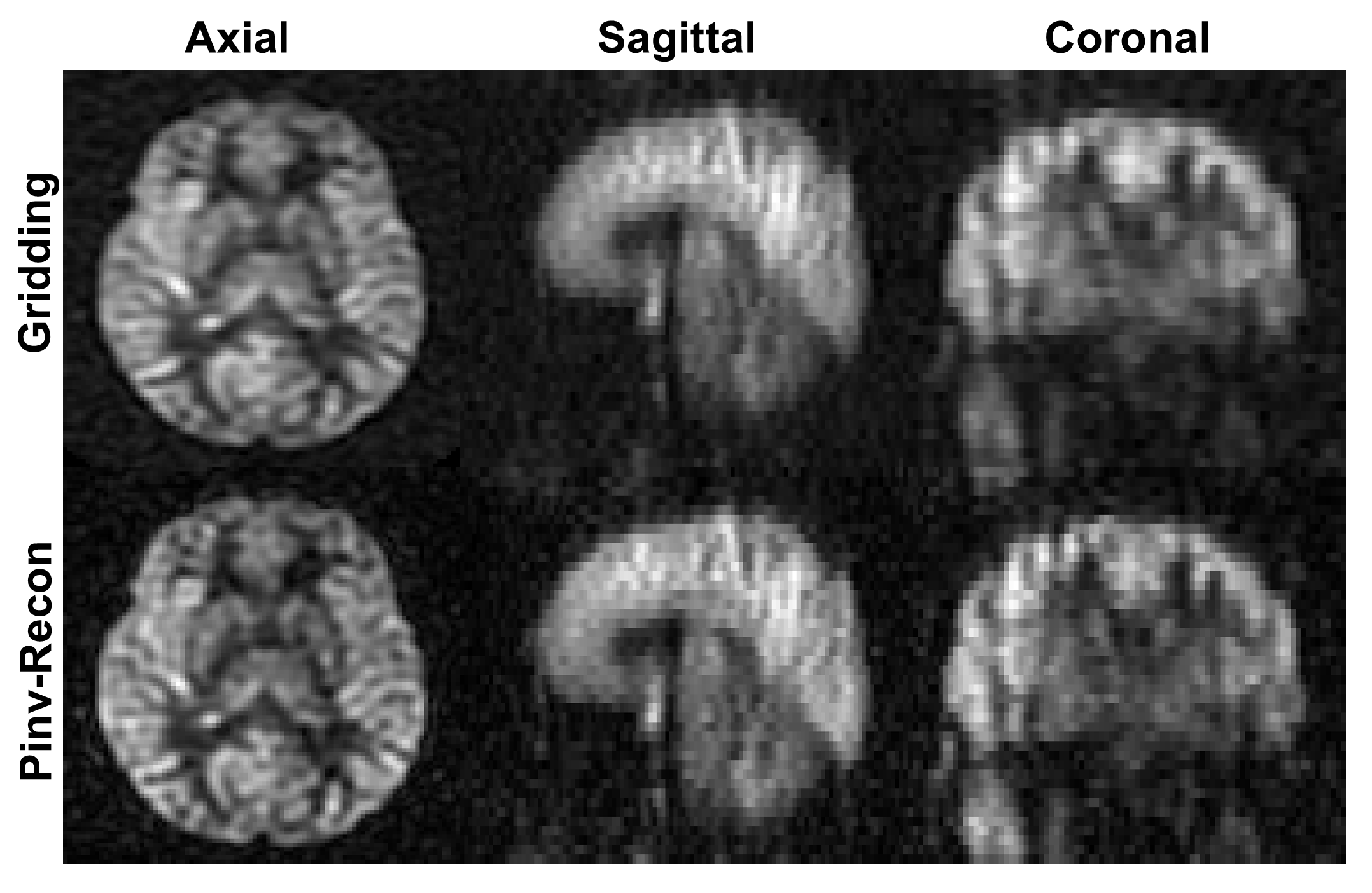}
    \caption{Example of reconstructing ASL data using Griding Pinv-Recon.}
    \label{fig:supp:asl}
\end{figure}

\end{document}